\pdfoutput=1 
\documentclass[conference]{IEEEtran}
\IEEEoverridecommandlockouts
\usepackage{cite}
\usepackage{graphicx}
\usepackage{balance}  
\usepackage{color}
\usepackage{amsfonts,amssymb,amsmath,bm}
\usepackage{makecell}
\usepackage{pifont}
\usepackage{stfloats}
\usepackage{subfigure}
\usepackage{multirow}
\usepackage[table,xcdraw]{xcolor}
\usepackage{graphicx}
\usepackage[ruled,vlined]{algorithm2e}
\usepackage{verbatim}
\usepackage{adjustbox}
\usepackage{makecell}
\usepackage{tabularx}
\usepackage{threeparttable}
\usepackage{url}

\usepackage{multirow}
\usepackage{graphicx}
\usepackage[normalem]{ulem}
\usepackage{graphicx}
\usepackage{ulem}
\usepackage{amsthm}
\useunder{\uline}{\ul}{}

\newtheorem{example}{Example}
\newtheorem{definition}{Definition}

\begin{document}


\title{GPU-Accelerated Batch-Dynamic \\Subgraph Matching}


\author{
Linshan Qiu$^{\dagger}$, Lu Chen$^{\dagger}$, Hailiang Jie$^{\dagger}$, Xiangyu Ke$^{\dagger}$, Yunjun Gao$^{\dagger}$, Yang Liu$^{\S}$, Zetao Zhang$^{\S}$\\

$^{\dagger}$\emph{Zhejiang University, Hangzhou, China} \qquad 
\normalsize $^{\S}$\emph{Huawei, Chengdu, China}\\

\emph{$^{\dagger}$\{lsqiu, luchen, hljie, xiangyu.ke, gaoyj\}@zju.edu.cn \qquad $^{\S}$\{liuyang169, zhangzetao3\}@huawei.com}  \\
}

\maketitle

\begin{abstract}
Subgraph matching has garnered increasing attention for its diverse real-world applications. Given the dynamic nature of real-world graphs, addressing evolving scenarios without incurring prohibitive overheads has been a focus of research.
However, existing approaches for dynamic subgraph matching often proceed serially, retrieving incremental matches for each updated edge individually. This approach falls short when handling batch data updates, leading to a decrease in system throughput.
Leveraging the parallel processing power of GPUs, which can execute a massive number of cores simultaneously, has been widely recognized for performance acceleration in various domains. Surprisingly, systematic exploration of subgraph matching in the context of batch-dynamic graphs, particularly on a GPU platform, remains untouched.

In this paper, we bridge this gap by introducing an efficient framework, \underline{\textsf{GAMMA}} (\underline{\textsf{G}}PU-\underline{\textsf{A}}ccelerated Batch-Dyna\underline{\textsf{m}}ic Subgraph \underline{\textsf{Ma}}tching). Our approach features a DFS-based warp-centric batch-dynamic subgraph matching algorithm. To ensure load balance in the DFS-based search, we propose warp-level work stealing via shared memory. Additionally, we introduce coalesced search to reduce redundant computations.
{Comprehensive experiments demonstrate the superior performance of \textsf{GAMMA}. Compared to state-of-the-art algorithms, \textsf{GAMMA} showcases a performance improvement up to hundreds of times.}


\end{abstract}

\vspace{-1mm}
\section{Introduction}
\vspace{-1mm}

\begin{figure}[t]
  \centering
  \vspace{-1mm}
  \includegraphics[width=\linewidth]{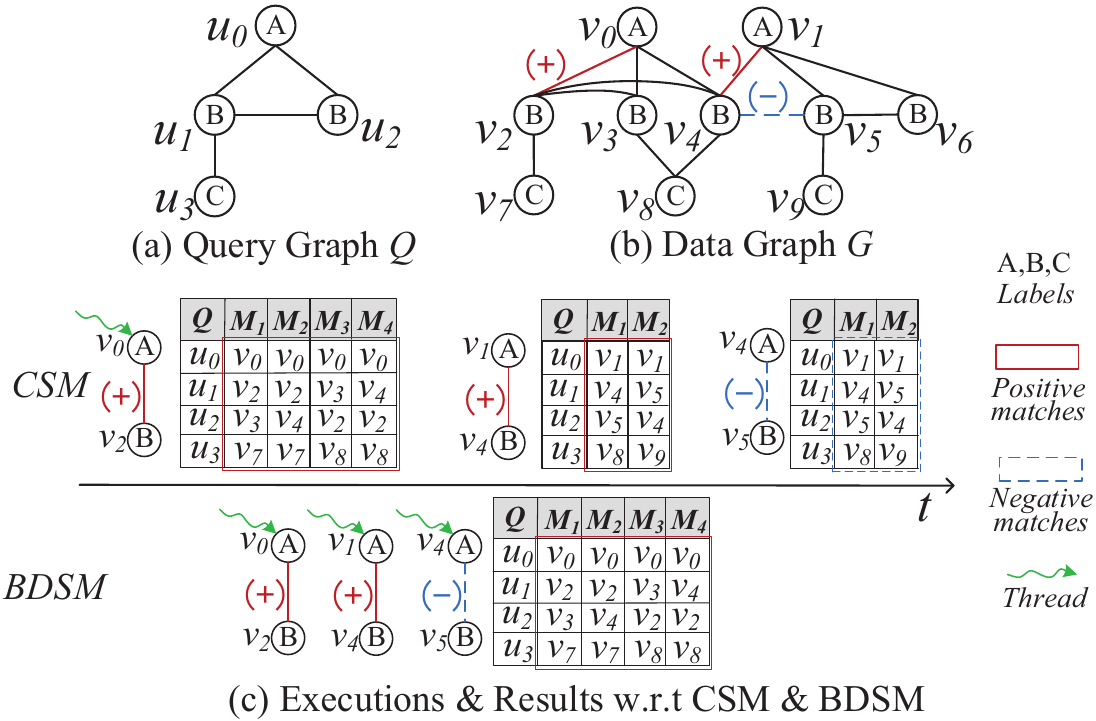}
  \vspace{-8mm}
  \caption{{A running example of batch-dynamic subgraph matching}}
  \label{fig:example}
  \vspace{-8mm}
\end{figure}

{Subgraph matching involves identifying subgraphs in a given data graph that are isomorphic to a targeted query graph. As an example, in Figure~\ref{fig:example}, $\{(u_0,v_1),(u_1,v_5),(u_2,v_6)$, $(u_3,v_9)\}$ is a match of query graph $Q$ in data graph $G$.} This process plays a crucial role in various domains~\cite{10.1145/2463676.2465300,sun2020memory,besta2022motif}, including network alignment, graph learning, and VLSI placement, among others. For instance, in VLSI placement, engineers leverage subgraph matching to pinpoint and replace areas that can be optimized, thereby mitigating costs.

Considerable efforts have been directed towards the development of algorithms tailored for static scenarios~\cite{1323804,ullmann1976algorithm,10.14778/1453856.1453899,10.1145/1516360.1516384,DBLP:journals/pvldb/ZhaoH10,10.1145/3299869.3300086,10.1145/2882903.2915236,10.1145/2463676.2465300,10.1145/1376616.1376660,jamshidi2020peregrine,10.1145/3469379.3469383}. However, real-life applications often involve evolving graph structures. Simply applying the aforementioned static approaches to dynamic scenarios necessitates {\em re-finding matches from scratch}, leading to prohibitively high computational overhead. More specifically, they are required to compute matches resulting from the consecutive updates of two edges and identify the incremental matches. To address these challenges, researchers have explored the continuous subgraph matching (CSM)~\cite{10.14778/3551793.3551803,10.1145/3183713.3196917,10.14778/3457390.3457395,10.14778/3342263.3342643,kankanamge2017graphflow,choudhury2015selectivity,10.1145/2489791}, which searches for incremental matches for each individual update in a {\em sequential} manner. 


{In the current landscape of data processing, characterized by immense dataset sizes and rapid transformations, updates are often grouped and applied as {\em batches}~\cite{dhulipala2020parallel}. This shift has led to a recent surge of interest in the development of efficient {\em parallel} {\em batch-dynamic} algorithms~\cite{10.1145/3514221.3517883,10.1145/3323165.3323196,9458741,liu2022parallel,dhulipala2020parallel}. Building on this trend, our paper delves into the intricate problem of subgraph matching under a batch-dynamic setting (BDSM). In BDSM, incremental matches are computed for a batch of updates rather than handling them individually.} 
{This problem holds practical significance in various domains, including cellular networks~\cite{iyer2015celliq}, social networks~\cite{chen2016realtime}, and e-commerce platforms~\cite{9458741}. In these scenarios, graph databases are collected and updated in batches, leveraging subgraph matching for tasks such as identifying patterns of malicious activity~\cite{gao2016toward}.} Despite the extensive literature on dynamic graph subgraph matching, i.e., the aforementioned CSM, scant attention has been given to batch-dynamic subgraph matching algorithms that enable parallel processing of update batches. 
{To illustrate this distinction, a comparative example between CSM and BDSM is presented below.}

\begin{example}
\vspace{-1.2mm}
{
In Figure~\ref{fig:example}(b), three updates $(v_0,v_2)$, $(v_1,v_4)$, $(v_4,v_5)$ unfold sequentially, with insertions denoted by $(+)$ and deletions by $(-)$. The detailed execution is illustrated in Figure~\ref{fig:example}(c), comparing CSM and DBSM. CSM processes updates individually, starting with $(v_0,v_2)$, generating four affirmative matches. Subsequent updates $(v_1,v_4)$ and $(v_4,v_5)$ result in two positive and two negative matches, respectively. Conversely, BDSM tackles these updates collectively as a parallel batch, yielding four positive matches. Note that BDSM disregards the order of updates, focusing solely on the matches post-batch update~\cite{9458741}, hence eliminating the redundant matches concerning $(v_1, v_4)$ and $(v_4, v_5)$ by CSM.
In conclusion, BDSM \textbf{(\romannumeral1)} 
mitigates the computational load by consolidating updates into batches, and \textbf{(\romannumeral2)} lends itself to facile parallelization as each update is independent.
}
\end{example}


{With the development of modern hardware boasting high computational capacities~\cite{10.1145/3128571,10.1145/3555041.3590816,DBLP:journals/pvldb/XekalakiFSDKBKK22}, there has been a dedicated research effort to harness these advancements in addressing traditional computation-heavy tasks~\cite{8547581,8000612,280876,10.14778/3389133.3389137,10.14778/3357377.3357379}.} 
Among these hardware options, GPUs, with their numerous computing cores~\cite{10.1145/3128571}, show great promise. 
{Surprisingly, systematic exploration of subgraph matching in the context of batch-dynamic graphs, particularly on a GPU platform, remains largely untouched.} 
{Enabling each update to be independently processed, yet retaining a comprehensive view of all updates within a batch, harnesses the parallel processing capabilities of GPUs to effectively address BDSM. The intricacies of the subgraph matching process involve frequent yet straightforward operations, like set intersections, contributing significantly to its exponential algorithmic complexity\footnote{{Set intersections take 58.2\% of total runtime in subgraph matching~\cite{han2018speeding}.}}. 
Leveraging the multitude of simpler cores in GPUs proves highly conducive to efficiently managing these computations, aligning well with the demands of the problem at hand.}
The primary challenges stem from the disparate architectures of CPUs and GPUs, characterized by variations in core placement and memory hierarchy. These differences result in {\em distinct execution modes and memory access patterns}. To ensure the optimal execution of algorithms on GPUs, it is imperative to meticulously devise algorithms tailored to exploit the unique strengths of GPU architectures. Given the nature of our problem and the underlying hardware architecture, three key challenges come to the forefront.

\textit{{\bf \em Challenge \uppercase\expandafter{\romannumeral1}:} How to maximize the GPU resource utilization for enhanced parallelism?} 
GPUs, with their hierarchical composition of threads and memory, feature a relatively smaller memory capacity but a significantly larger number of cores compared to CPUs~\cite{10.1145/3128571}.
Existing graph processing algorithms~\cite{10.1145/1516360.1516384,kankanamge2017graphflow,10.1145/2489791} often resort to exploration based on Breadth-First Search (BFS) to exploit parallelization capabilities, albeit incurring substantial {\em memory overhead}.
{In contrast, exploration based on Depth-First Search (DFS) significantly reduces memory consumption by exploring only a small portion of vertices at a time~\cite{280876}. 
However, DFS introduces challenges related to {\em load imbalance} due to the unpredictable workload associated with each task. The need to retrieve multi-hop neighbors during DFS search, coupled with significant variations in adjacency lists, exacerbates the issue.}
Given the constraints of GPU memory, our system adopts DFS search for match retrieval. To address the associated performance challenges, we establish a baseline with a warp-centric granularity (\S\ref{sec:gpu_dfs}).
This approach effectively mitigates load imbalance by leveraging the memory hierarchy of GPUs. Specifically, we assign a warp to cooperatively handle an updated edge, thereby reducing thread divergence.
Furthermore, the threads within a warp remain together to facilitate memory transitions in a coalesced manner, leading to a substantial reduction in memory divergence. 
To tackle load imbalance, we propose a work-stealing strategy, distributing workload information for each warp in shared memory. This enables idle warps to seize work from active ones, significantly enhancing thread utilization and overall system efficiency (\S\ref{sec:handling_load_imbalance}).

\textit{{\bf \em Challenge \uppercase\expandafter{\romannumeral2}:} How to avoid unnecessary graph traversing to improve the matching efficiency?} In BDSM, inherent computational redundancies during the search primarily arise from {\em automorphisms} present in query subgraphs. 
{A vivid example is illustrated with the insertion of $e(v_0,v_3)$ in $G$ (Figure~\ref{fig:example}). When mapping $e(v_0,v_2)$ to $e(u_0,u_1)$, the incremental matches are $M_1$ and $M_2$.
Similarly, mapping this update to $(u_0,u_2)$ leads to additional matches $M_3$ and $M_4$.
Remarkably, the subgraph induced by vertices ${v_0,v_2,v_3}$ is visited twice during this process. This redundancy emanates from the automorphisms present in the subgraph of the query graph, specifically the induced subgraph composed of ${ u_0,u_1,u_2 }$.} 
To mitigate this redundancy, we introduce the coalesced search technique {(\S\ref{sec:cs})}, anchored in the concept of $k$-degenerated automorphic subgraph. This subgraph retains its automorphisms even after removing $k$ vertices from the original graph.
Using the $k$-degenerated automorphic subgraph, we aggregate equivalent edges into a group. Leveraging the $k$-degenerated automorphic subgraph, we aggregate equivalent edges into a group. Hence, we need only consider one edge among the equivalent edges during mapping and generate other partial matches through permutation operations. This innovative approach significantly reduces the redundant traversals, thereby streamlining the matching process and improving overall efficiency.

\textit{{\bf \em Challenge \uppercase\expandafter{\romannumeral3}:} How to harmonize modules to achieve efficient computation pipelining?} This challenge is crucial from a system development perspective, requiring meticulous groundwork in preprocessing and graph updating to facilitate incremental subgraph matching at higher levels and achieve optimal performance. 
Our objective is to pave the path for incremental subgraph matching while minimizing the time overhead of preceding steps prior to result computation. To achieve this, we adopt an asynchronous approach {(\S\ref{sec:overview})}, conducting preprocessing on the CPU concurrently with GPU computations to alleviate waiting times. Facing the continuous arrival of update batches, efficient application of updates to the data graph becomes paramount. 
Inspired by this, we embrace the widely used GPMA~\cite{sha2017accelerating} as the underlying dynamic graph structure. However, certain inefficiencies arise when dealing with small-sized segments and locating segments with GPMA. Consequently, we introduce practical optimizations by leveraging the Cooperative Group and shared memory, respectively, making the entire computational pipeline more seamless and responsive to the continuous influx of update batches {(\S\ref{sec:graph_container})}.

To sum up, the key contributions are summarized as follows:

\begin{itemize}
\item We introduce \textsf{GAMMA}, the first GPU-based approach tailored for efficient batch-dynamic subgraph matching. This groundbreaking proposal leverages the parallel processing power of GPUs, marking a significant advancement in this domain.

\item We design a warp-centric batch-dynamic subgraph matching algorithm that capitalizes on GPU parallelism, addressing challenges like thread and memory divergence. Additionally, we introduce a warp-centric work-stealing strategy, utilizing shared memory to balance workloads among warps within the same block.

\item {We devise a coalesced search technique to tackle computational redundancies arising from subgraph automorphisms. Unnecessary computations are minimized during matching by only finding matches for one automorphism and deriving other partial ones through permutations.}


\item We conduct extensive experiments on 6 public datasets, showcasing that our framework significantly outperforms popular advanced methods by up to hundreds of times. 
\end{itemize}  

The rest of the paper is organized as follows. We formulate our problem and introduce GPUs in Section~\ref{sec:preli}. Section~\ref{subsec:relate_work} reviews the related work. Sections~\ref{sec:method} and ~\ref{sec:opt} illustrate our designs. The experimental results and our findings are reported in Section~\ref{sec:experiments}. Finally, Section~\ref{sec:conclusion} concludes the paper.

\section{Preliminaries}

In this section, we formally define the problem and introduce GPUs. The frequently used notations are listed in Table~\ref{tab:symbol}.
\label{sec:preli}

\vspace{-5mm}
\subsection{Problem Definition}

We first introduce some basic concepts, and then we formulate the studied problem. 

Let $g=(V,E,L)$ be an undirected labeled graph, where $V$ denotes the vertex set, $E \subseteq V\times V$ is the edge set, and $L$ is a mapping function that maps a vertex $v \in V$ or an edge $e \in E$ to a label $l$ in a label set $\Sigma$. {Take the query graph $Q$ in Figure~\ref{fig:example}(a) for example. The label set of $Q$ is $\Sigma = $ \{A, B, C\}, and the label of $u_0$ is A, i.e., $L(u_0) = $ A. Specifically, in biological networks, vertices are labeled with protein types, and then, biologists can count the occurrences of a particular pattern to determine the property. In recommendation systems, users are labeled with different attributes, and then, a company can find its target customers by detecting community patterns.} Here, we use $e(u,u^\prime)$ to represent the edge between vertices $u$ and $u^\prime$. Given a vertex $v \in V$, we denote its neighbor set as $N(v)$ and $deg(v) = |N(v)|$ as its degree. Besides, $N^l(v)$ are the neighbors of $v$ that have the label $l$. In our problem, two types of graphs are involved, i.e., query graph $Q$ and data graph $G$. We call vertices and edges in the query graph $Q$ (resp. data graph $G$) query vertices and query edges (resp. data vertices and data edges). 

\begin{table}
\centering
    \small
    \caption{Symbols and Descriptions}
    \vspace{-3mm}
    \label{tab:symbol}
    \setlength{\tabcolsep}{3pt}
    \begin{tabular}{|c|c|}
    \hline
    \textbf{Symbos} & \textbf{Descriptions} \\
    \hline\hline
        $G$, $Q$ & the data graph and the query graph\\ \hline
        $V(Q)$, $E(Q)$ & the vertex set and the edge set of $Q$   \\ \hline
        $e(u,u^\prime)$ & an edge between vertex $u$ and $u^\prime$\\ \hline
        $L(u)$ & the label of vertex $u$ \\ \hline
        $N(u), N^{l}(u)$ & neighbors and neighbors with label $l$ of vertex $u$\\ \hline
        $deg(u)$ & the degree of vertex $u$ \\ \hline
        $\Delta\mathcal{G}, \Delta G, \Delta e$ & graph stream, graph update, single update \\ \hline
        $\oplus = +/-$ & an edge insertion/deletion \\ \hline
        $C(u)$ & the candidate set of data vertex $u$\\ \hline
        $\Delta \mathcal{M}$ & incremental matches \\ \hline
        $\pi$ & matching order \\ \hline
    \end{tabular}
\vspace{-6mm}
\end{table}

\begin{definition}
\label{def:dynamic}
A graph update stream $\Delta \mathcal{B} = (\Delta B_1, \Delta B_2,\dots )$ is a sequence of update operations, where $\Delta B$ is a set of edge insertions/deletions $\{\Delta e_i\}$, as $\Delta e_i = (\oplus,e_i)$ is the insertion/deletion of an edge $e_i$, where $\oplus$ denotes insertion $(+)$ or deletion $(-)$. 
\end{definition}

A graph is batch-dynamic when each update stream contains multiple insertion/deletion operations, that is $|\Delta B| > 1$. Note that, vertex insertions/deletions can be treated as edge insertions/deletions. For vertex insertions, we first insert the vertices into the data graph and regard the incident edges as a collection of edge insertions. For vertex deletions, we first remove the corresponding vertices and treat the incident edges as a collection of edge deletions.

\begin{definition}
Given a query graph $Q$ and a data graph $G$, a subgraph isomorphism of $Q$ in $G$ is a bijective function $M$ between $V(Q)$ and $V(G_s)$, where $G_s$ is a subgraph of $G$ such that (1) $\forall u \in V(Q), L(u)=L(M(u))$; and (2) $\forall e(u, u^{\prime}) \in E(Q), e\left(M(u), M(u^{\prime})\right) \in E\left(G_s\right)$.
\label{def:iso}
\end{definition}

According to Definition~\ref{def:iso}, we call a subgraph isomorphism an embedding or a match. As the example shown in Figure~\ref{fig:example}, a subgraph isomorphism of $Q$ in $G$ is {$M = \{ M(u_0) = v_1, M(u_1) = v_5, M(u_2) = v_6, M(u_3) = v_9\}$}. Given a graph update $\Delta B\in \Delta \mathcal{B}$, we denote $G^\prime$ the graph resulted from applying $\Delta B$ to $G$. The incremental matches $\Delta \mathcal{M}$ with respect to the graph update $\Delta B$ is the difference between $\Delta \mathcal{M}$ and $\Delta \mathcal{M}^\prime$, where $\Delta \mathcal{M}$ and $\Delta \mathcal{M}^\prime$ represent the matches in $G$ and $G^\prime$, respectively. 

Existing continuous subgraph matching (CSM) algorithms~\cite{10.14778/3551793.3551803,10.1145/3183713.3196917,10.14778/3457390.3457395,10.14778/3342263.3342643,kankanamge2017graphflow,choudhury2015selectivity,10.1145/2489791} assume that the graph update contains only a single edge update, i.e., $|\Delta B| = 1$, and aim to find the incremental matches $\Delta \mathcal{M}$ for each single updated edge separately. To fully utilize the power of parallelism a GPU provides, we suppose $|\Delta B| > 1$ and perform batch-dynamic subgraph matching on $\Delta B$.

\textbf{Problem Statement.}
Given update batches $\Delta \mathcal{B}$, batch-dynamic subgraph matching (\textbf{BDSM} for short) finds all incremental matches $\Delta \mathcal{M}$ for each $\Delta B \in \Delta \mathcal{B}$, where $|\Delta B > 1|$.


\subsection{Graphics Processing Units (GPUs)}
\begin{figure}[tb]
    \centering
    \includegraphics[width=0.31\textwidth]{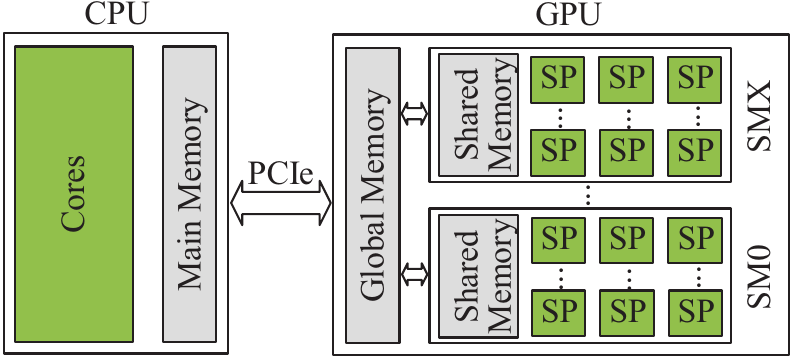}
    \vspace{-3mm}
    \caption{The simplified GPU architecture}
    \vspace{-6mm}
    \label{fig:cpugpu}
\end{figure}

{In contrast to CPUs, which have only a few cores, GPUs offer thousands of lightweight cores. CUDA (Compute Unified Device Architecture) provides a programming abstraction that connects applications with GPU hardware. Refer to Figure~\ref{fig:cpugpu} for a simplified GPU architecture.}

{
\textbf{Thread Hierarchy.} A modern GPU consists of streaming multiprocessors (SMX), each containing streaming processors (SP). When a program (known as a kernel in CUDA), multiple threads collaborate to perform computations. Threads use SPs, with 32 threads forming a warp. Warps are the SM's scheduling units, operating in a single-instruction multiple-thread (SIMT) manner. Threads that meet a branch condition execute concurrently, while others idle, a situation called warp divergence to avoid. Several warps create a Cooperative Thread Array (CTA) or block, assigned to an SM and unchangeable during runtime. Blocks form a grid, encompassing all GPU threads.
}

{
\textbf{Memory Hierarchy.} GPUs are memory-efficient due to their high-bandwidth and parallel access design. They feature multiple memory levels: global memory (shared by all threads, with slower access), shared memory (small but with high bandwidth, allocated to thread blocks), and registers (smallest, for frequently accessed data). To improve efficiency, threads within the same warp should aim for consecutive memory accesses, called memory coalescing. Non-consecutive access results in lower bandwidth use, underscoring the importance of data locality. 
}

\section{Related Work}
\label{subsec:relate_work}

We briefly review related work about subgraph matching on static graphs and dynamic counterparts, followed by GPU-accelerated graph processing.

\subsection{Subgraph Matching}
{
Subgraph matching, the task of identifying query graph $Q$ occurrences within data graph $G$, has seen extensive research.  Ullmann~\cite{ullmann1976algorithm} introduces a backtracking algorithm that expands a partial embedding by sequentially mapping query vertices to data vertices. Subsequent research falls into three categories: the first ~\cite{10.14778/1453856.1453899,1323804} directly searches the data graph to find matches, suffering from prohibitive space and computational overheads. The second ~\cite{10.1145/1516360.1516384,DBLP:journals/pvldb/ZhaoH10} constructs an index for the data graph before searching. The third category~\cite{10.1145/1376616.1376660,10.1145/2463676.2465300,10.1145/2882903.2915236,10.1145/3299869.3300086} creates candidate sets for query vertices, maintains edges between them using an auxiliary data structure. Subsequently, a matching order is generated, and the matched results are enumerated.
}

{
However, the aforementioned approaches remain largely oblivious to the query graph's structure, potentially exploring unnecessary subgraphs. Recently, there have been advancements in pattern-aware subgraph matching algorithms~\cite{jamshidi2020peregrine,10.1145/3469379.3469383}. Yet, applying such static subgraph matching algorithms to dynamic scenarios necessitates rebuilding the index and searching the differential matched subgraphs between snapshots, incurring excessive space and computational costs.
}

\subsection{Continuous Subgraph Matching}

{
Numerous studies have addressed efficient subgraph matching in dynamic graphs(CSM)~\cite{10.14778/3551793.3551803,10.1145/3183713.3196917,10.14778/3457390.3457395,10.14778/3342263.3342643,kankanamge2017graphflow,choudhury2015selectivity,10.1145/2489791,yang2023fast}. IncIsoMat~\cite{10.1145/2489791} extracts relevant subgraphs from the data graph and performs subgraph matching before and after updates. However, it enumerates unnecessary matches, leading to substantial computational overhead. Recent approaches use an incremental style for CSM. For example, Graphflow~\cite{kankanamge2017graphflow} maps updated edges to the query graph and extends partial results by repeatedly joining the remaining vertex of the query graph. SJ-Tree~\cite{choudhury2015selectivity} employs index-based binary joins but requires significant memory storage. TurboFlux~\cite{10.1145/3183713.3196917} efficiently solves CSM using a data-centric graph (DCG). SymBi~\cite{10.14778/3457390.3457395} maintains a directed acyclic graph and embeds weak embeddings of directed acyclic graphs to quickly retrieve matches and support efficient updates. RapidFlow~\cite{10.14778/3551793.3551803} reduces CSM to batch subgraph matching (BSM), upon which an effective matching order can be generated. Calig~\cite{yang2023fast} minimizes incremental match generation time by reducing backtracking.
}
 
{
Despite reducing recomputation, these methods conduct CSM sequentially. Motivated by this, we investigate batch-dynamic subgraph matching to enhance efficiency. Batch processing excels at handling substantial volumes of evolving data. Several algorithms have been developed for batched-updates, including computing clustering coefficients~\cite{10.1145/1292609.1292616}, single-source shortest-path~\cite{10.1145/1516360.1516384}, dynamic connectivity problems~\cite{10.1145/2882903.2915236}, and triangle counting~\cite{10.1145/2882903.2915236}. However, subgraph matching in batch-dynamic graphs, a fundamental problem in graph processing, remains unexplored to the best of our knowledge.
}

\subsection{GPU-accelerated Graph Processing}
\label{subsec:related_work_gpu}
{
The emergence of novel hardware, particularly GPUs, presents opportunities to expedite a range of computational tasks~\cite{10.1145/3128571}, including graph processing tasks like clique counting~\cite{10.1145/3128571}, subgraph enumeration~\cite{7468500,10.1145/3318464.3389699}, pattern mining~\cite{280876,10.14778/3389133.3389137}, and PageRank computation~\cite{10.14778/3357377.3357379}. Systems like Gunrock~\cite{10.1145/3108140}, Pangolin~\cite{10.14778/3389133.3389137}, and GraphPEG~\cite{10.1145/3450440} simplify graph analysis on GPUs. However, these tools primarily target static graphs. To hasten dynamic graph processing, GPMA~\cite{sha2017accelerating} employs the Packed Memory Array (PMA) for quick updates. Subsequently, cuSTINGER~\cite{green2016custinger}, aimGraph~\cite{winter2017autonomous}, FaimGraph~\cite{winter2018faimgraph}, Hornet~\cite{busato2018hornet}, and others have proposed various data structures for dynamic graphs. These data structures are fundamentally intended to curtail dynamic graph updates rather than diverse graph processing tasks. This paper investigates GPUs-accelerated batch-dynamic subgraph matching.
}

\section{Method}
\label{sec:method}

In this section, we will initially furnish an overview of our method, followed by a detailed exposition of each component.

\subsection{Overview}
\label{sec:overview}

Figure~\ref{fig:framework} provides an insightful overview of our innovative CPU-GPU heterogeneous framework, \textsf{GAMMA}, for batch-dynamic subgraph matching. The framework comprises four key components: Preprocess, Update, Computational Kernel (BDSM), and Postprocess. 
For each batch, the preprocess component conducts fundamental analyses, such as {maintaining neighborhood information and generating candidate sets. These outcomes, along with the updated edges, are then dispatched to the GPU.  
Subsequently, \textsf{GAMMA} executes the update on the data graph, followed by the computational kernel identifying incremental matches. 
Finally, the postprocess component utilizes the matching results for application-specific tasks. 

The computational kernel integrates two practical optimizations: work stealing (\S~\ref{sec:handling_load_imbalance}) and coalesced search (\S~\ref{sec:cs}). These optimizations strategically balance the workload and eliminate redundant computations. 
Importantly, all four components operate asynchronously. The computational kernel is intricately designed to overlap the preprocessing step and the host-to-device data transfer for the next batch. Likewise, once the matching results are generated, they seamlessly overlap with the next update and computation step. This asynchronous process continuously repeats, enhancing the overall efficiency.

\begin{figure}
  \centering
  \includegraphics[width=\linewidth]{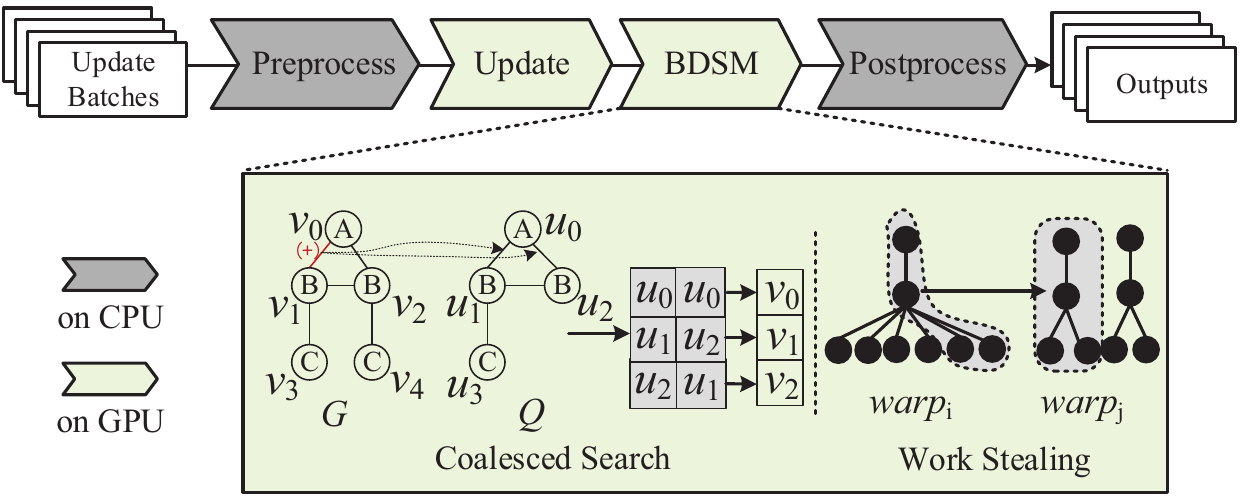}
  \vspace{-6mm}
  \caption{{Overview of \textsf{GAMMA}}}
  \label{fig:framework}
  \vspace{-6mm}
\end{figure}

\subsection{Preprocessing}

The primary function of preprocessing is to expeditiously and effectively generate candidate sets for each query vertex in the query graph. Existing studies have put forth a plethora of filtering, among which neighborhood-label-frequency-based~\cite{10.1145/2463676.2465300} are widely employed. 
The idea of this filtering strategy is that the data vertices in the candidate set for a query vertex should exhibit {\em a similar neighborhood structure}. To be more specific, for each candidate vertex $v$ in the candidate set $C(u)$ of a given query vertex $u$, in addition to the label constraint, i.e., $L(v)=L(u)$, each candidate data vertex $v \in C(u)$ should contain neighbors {with same labels from the query vertex's neighborhoods, and the number of such neighbors should not be fewer than the count of corresponding neighbors of the query vertex, i.e., $|N^l(v)| \geq |N^l(u)|$ where $l\in \{ L(u^\prime)| u^\prime \in N(u) \}$. 
Leveraging this concept, GSI~\cite{zeng2020gsi} applies binary encoding on vertices, where each vertex is represented by a $K$-bit binary code. Hence, the $K$-bit binary code is divided into two parts: the first $N$ bits for the vertex label, and the remaining $(K-N)$ bits for the neighborhood structure. The second part is further divided into groups of $M$ bits to record the neighbor counts with different labels. 
Inspired by GSI, we refine this encoding strategy to avoid encoding labels absent in the query graph. Figure~\ref{fig:gen_cand} illustrates the encoding results of $G$ before and after applying the updates (denoted as $G_1$ and $G_2$, respectively) and $Q$ in Figure~\ref{fig:example}}. 
{Regarding the encoding table of $G_2$, there are $K=9$ bits for each vertex and the first $N=3$ bits denote the label, e.g., ``001`` for label ``A'' on $v_0$ and $v_1$. The remaining $(K-N)=6$ bits, grouped in sets of $M=2$, depict the count of neighbors with specific labels. For instance, $v_0$ has three neighbors with label $B$, hence in the encoding table corresponding to the columns for label $B$, we set it as ``11''.}
The encoding is conducted on the CPU. To generate the candidate sets for each query vertex, we can simply perform a bitwise AND operation. For example, to generate the candidate set for query vertex $u$, we perform ``$ENC(u) \bigwedge ENC(v)$" for each data vertex $v$, {where $ENC(v)$ represents the encoding of $v$}. If the result is $ENC(u)$, then $v$ is a candidate of $u$. With high parallelism provided by GPU, the bitwise AND operation can be efficiently performed.

\begin{figure}[tb]
    \centering
    \includegraphics[width=0.48\textwidth]{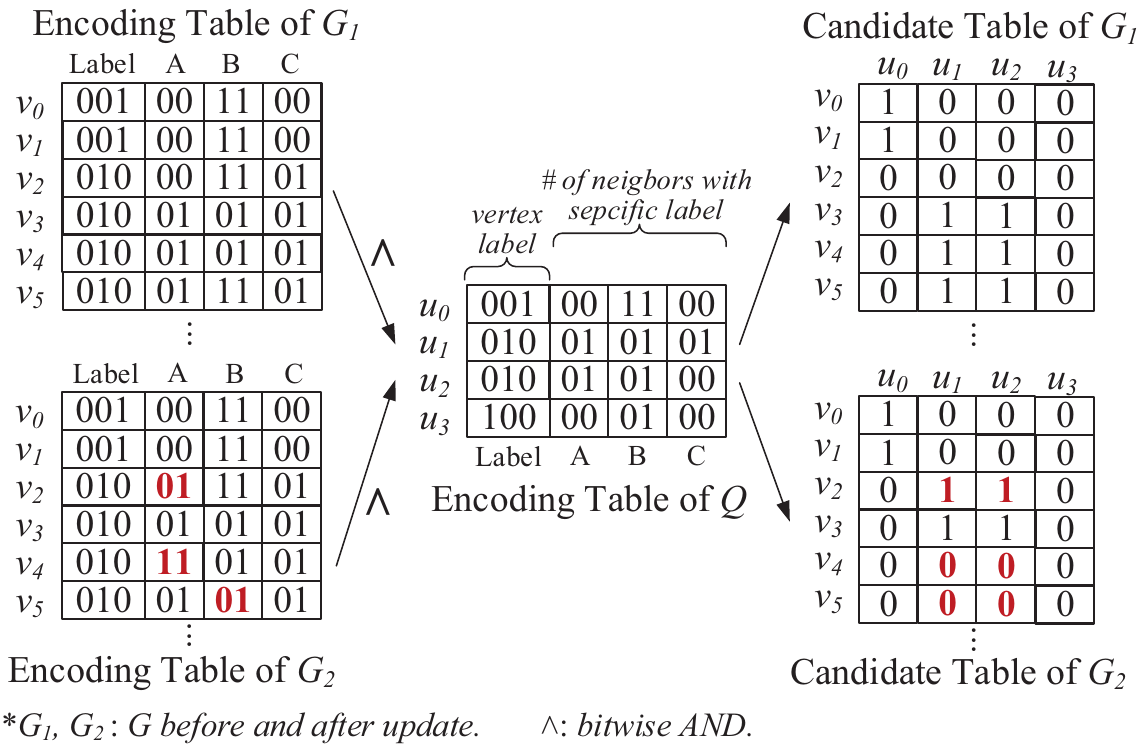} 
    \vspace{-4mm}
    \caption{{Preprocessing for candidate table generation. $v_6\sim v_9$ are omitted for brevity. In the example, we use the first 3 bits for vertex label encoding and the remaining 6 bits for counting neighbors with specific labels.}} 
    \vspace{-3mm} 
    \label{fig:gen_cand}
\end{figure}

\textbf{Encoding of dynamic graphs.} Given the dynamic nature of our problem setting, re-encoding data vertices for each batch is not only time-consuming but also incurs a substantial data transfer cost, potentially becoming a bottleneck. 
Henceforth, during the initialization phase, we compute the encodings for all data vertices. For subsequent batches, we load only the vertices with modified encodings into the GPU and compute corresponding candidate sets.  
Take $G$ in Figure~\ref{fig:example}(b) as an example. {After
applying the updates, the encodings of the corresponding vertices update accordingly, as depicted by the red bold font in the bottom left part of Figure~\ref{fig:gen_cand}. Specifically, following the insertion of $e(v_0,v_2)$, $v_2$ gains a neighbor with label $A$, prompting an increment from ``00" to ``01" in the respective column of the encoding table for $G_2$, marked in red. Notably, despite {$e(v_0,v_2)$} involves vertex $v_0$, its encoding remains unchanged due to our use of a 2-bit binary code, a trade-off between space and filtering capabilities. 
Subsequently, we compute candidate sets for query vertices, updating the candidate table (to be explained later) for $G_2$. For example, due to the insertion of {$e(v_0,v_2)$}, {$v_2$} will match $u_1$ and $u_2$.}

\textbf{Candidate Table.} Directly allocating an array for each query vertex to store its candidate data vertices incurs a substantial memory storage cost. Given the limited GPU memory, we introduce a more space-efficient representation known as the candidate table.
The structure of the candidate table is depicted in the right part of Figure~\ref{fig:gen_cand}, with each row corresponding to a data vertex and each column corresponding to a query vertex. Binary markers are employed to signify whether a data vertex belongs to the candidate set of a specific query vertex. 
For example, $v_0$ is a candidate vertex of $u_0$, and thus, we mark the entry in the first row and first column as 1 in the candidate table of $G_2$.

\subsection{Search Strategies {in BDSM Computational Kernel}}
\label{sec:gpu_dfs}
The update component can be seamlessly accommodated in our data structure proposed in \S\ref{sec:data_structure}.
In this subsection, we proceed to discuss the search strategy employed in \textsf{GAMMA}. 

{
\textbf{BFS vs DFS.} Current GPU-centric pattern mining algorithms predominantly optimize data locality and enhance parallelism by leveraging BFS{~\cite{280876}}.  
However, this approach often leads to an exponential increase in intermediate results, consuming a significant amount of memory~\cite{280876}. When device memory reaches its limit, the system resorts to moving data between system memory and global memory, causing further performance degradation~\cite{10.1145/3318464.3389699}. 
On the other hand, traditional CPU-based subgraph matching algorithms commonly adopt the DFS-based backtracking framework introduced by Ullmann~\cite{ullmann1976algorithm}. 
DFS-based searches significantly reduce memory overhead by materializing only the final results and avoiding the storage of numerous (invalid) partial matches.}

{
In Figure~\ref{fig:bfs_vs_dfs}(a), we illustrate the trends in device memory usage for BFS and DFS using the \textit{LS} dataset. It is apparent that BFS leads to exponential memory growth, quickly depleting the available memory. In contrast, DFS maintains a more gradual memory consumption profile, significantly lower than that of BFS. Due to memory exhaustion, a substantial volume of data transfers (Comm. in Figure~\ref{fig:bfs_vs_dfs}(b)) between main memory and device memory becomes necessary, exerting a dominant influence on the total time, even surpassing the computation time (Comp. in Figure~\ref{fig:bfs_vs_dfs}(b)) by several times. Additionally, the computational time of BFS also exceeds that of DFS, as BFS requires synchronization after each expansion.
}

Consequently, we opt for DFS to retrieve the incremental matches, prioritizing memory resources due to their significance and limited quantities. 
When performing DFS on CPUs, the number of simultaneously running threads is constrained by core limitations. 
{In the context of GPU execution, a massive number of threads operating concurrently exacerbate the load imbalance issue.} {Moreover, the task of estimating workloads and consumed memory in the DFS process proves to be a formidable challenge}. {Finally, the random access nature of DFS conflicts with the coalesced memory access characteristics of GPUs.} Therefore, the organization of threads becomes crucial.

\textbf{Choice of Thread Granularity}. In our problem setting, a task is responsible for searching the matches of an updated edge. Adopting various thread granularities may lead to notable performance disparities. The following discussion delves into the thread granularity choices for our specific problem.
\begin{itemize}
    \item \textbf{Thread-centric Assignment}. 
    Assigning a thread per update seems intuitive, offering high parallelism, but it introduces performance issues. A surplus of threads can lead to reduced resources per thread, and divergent branch conditions for different vertices result in thread divergence. Additionally, non-consecutive memory transactions for each thread cause memory divergence, incurring unnecessary costs for memory-fetching operations.
    \item \textbf{Warp-centric Assignment}. The warp-centric task assignment designates a warp to handle an update, enabling cooperative operations among its threads with low-cost primitives. 
    This helps address the memory divergence problem, as the 32 threads within a warp stay together during memory reading for adjacency lists. 
    Additionally, they collaborate on set intersections for extracting candidate sets, contributing to improved performance and reduced thread divergence. 
    Despite these benefits, workload imbalance remains inherent due to the unpredictable execution paths of DFS, and we will soon introduce a solution to address this issue among warps. (\S\ref{sec:handling_load_imbalance}).
    \item \textbf{Block-centric Assignment}. The block-centric strategy allocates more resources for each update. However, handling load imbalances between blocks requires high-latency device memory transactions. The limitation on the number of blocks that can execute concurrently in streaming multiprocessors presents a constraint that we acknowledge and plan to explore in future work.
\end{itemize}

\begin{figure}[tb]
    \centering
    \includegraphics[width=0.48\textwidth]{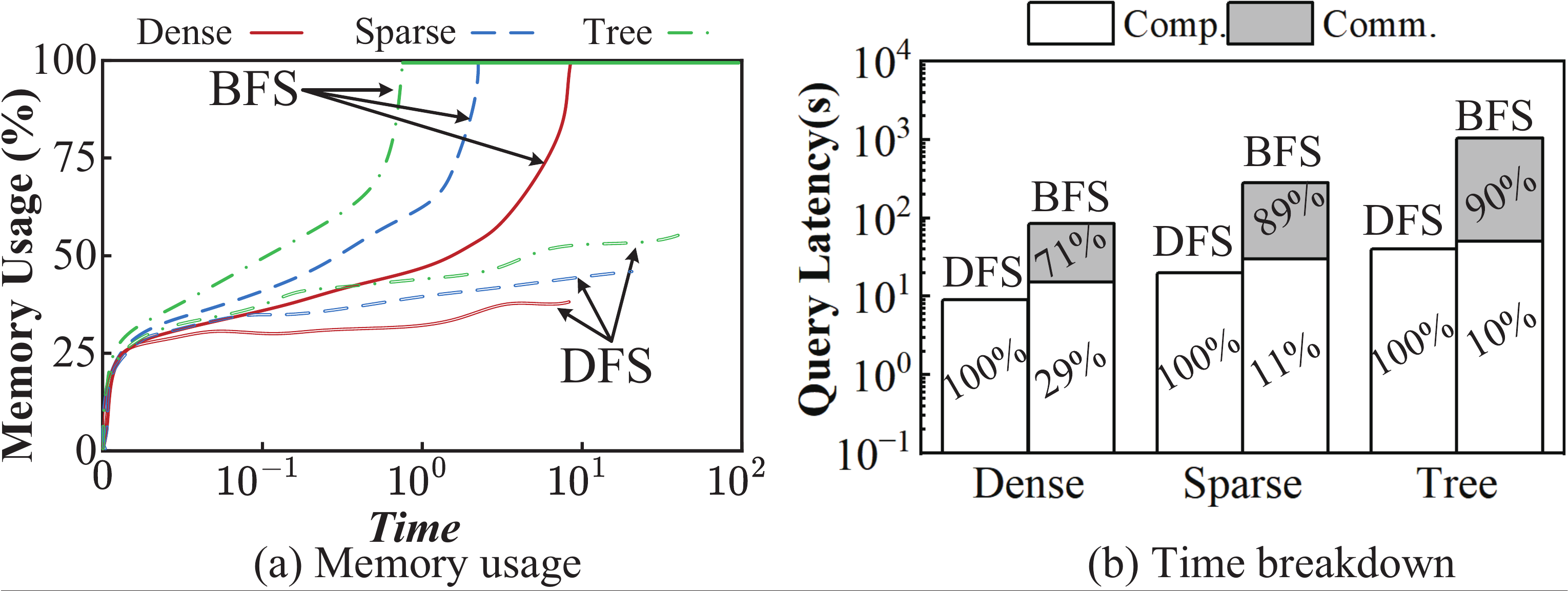} 
    \vspace{-3mm}
    \caption{{A comparison of BFS and DFS in a GPU environment}}
    \label{fig:bfs_vs_dfs}
\end{figure}

\setlength{\textfloatsep}{0pt}
\begin{algorithm}[t]
\small
\caption{Warp-centric batch-dynamic subgraph matching algorithm (WBM)}
\label{alg:alg1}
\LinesNumbered
  \KwIn{data graph $G$, query graph $Q$, update stream $\Delta\mathcal{B}$, matching order $\pi$ }
  \KwOut{incremental matches $\Delta\mathcal{M}$}
  \While{ $\Delta B \subseteq \Delta \mathcal{B} \neq null$ }{
    \ForEach{ $\Delta e \in \Delta B$}{
     \tcp{do in parallel}
      \ForEach{ $e(u_1, u_2) \in E(Q)$}{ 
         \If{$L(v_1) = L(u_1)$ \&\& $L(v_2)=L(u_2)$}{
            $M \leftarrow \{(u_1, v_1), (u_2, v_2)\}$\;
            $l=2$; ${p[l]=0}$\;
            $C[l]=\mathsf{GenCandidates}$($G$, $Q$, $M$, $\pi$, $l$)\;
            
            \While{$l \ge 2$ \&\& ${p[l]} < C[l].size$}{
                \eIf{$l=(|V(Q)|-1)$}{
                    \ForEach{$c \in C[l]$}{ 
                        $\Delta \mathcal{M}+=\{M, (\pi[l], c)\}$;
                    }

                    \While{$(--l\geq 2)$ \&\& $(++{p[l]}\geq C[l].size)$}{pop an entry from $M$\;}
                }{
                    \While{${p[l]}<C[l].size$}{
                        $C[l+1]$ = $\mathsf{GenCandidates}$($G$,$Q$,$M\cup (\pi[l],C[l][{p[l]}])$,$\pi$,$l+1$)\;

                        \If{$C[l+1] \neq \O$}{ $M += (\pi[l],C[l][{p[l]}])$\;$l++$,${p[l]=0}$, break;}
                        \lElse{${p[l]++}$}
                    }
                   
                    \If{${p[l]} \geq C[l].size$}{execute as lines 12--13\;}

                }
            }
        }
      }
    }
  }

{
\textbf{Procedure} $\mathsf{GenCandidates}(G,Q,M,\pi,l)$\\
$res=CTable[\pi[l]]$;\tcp{$CTable$: candidate table}
\For{$i=0$ $\mathrm{to}$ $(l-1)$}{
    \If{$\pi[i] \in N(\pi[l])$}{
        $v=M[\pi[i]]$\;
        \tcp{intersect $res$ with $N(v)$}
        $res = \mathsf{Intersection}(res,N(v))$; 
    }
}
\Return $res$
}
\end{algorithm}
\setlength{\textfloatsep}{0pt plus 2pt minus 2pt}

Algorithm~\ref{alg:alg1} depicts our warp-centric batch-dynamic subgraph matching algorithm (termed as WBM), i.e., assigning a warp to process each updated edge concurrently. WBM takes a data graph $G$, a query graph $Q$, update {stream} $\Delta\mathcal{B}$, and a matching order $\pi$ as inputs, and outputs the incremental matches $\Delta\mathcal{M}$. The matching order guides the order in which query vertices are matched, and we generate it for each query edge offline. {The matching order tends to prioritize the more selective query vertices, such as those with higher degrees and fewer candidates, which can provide more strict constraints for candidate generation and minimize the candidate size, hence pruning the search space}. 
Under the update stream $\Delta \mathcal{B}$, WBM conducts a while-loop to enumerate each update batch $\Delta B$ (lines 1--22). Hence, a for-loop processes each record of $\Delta B$ in parallel (lines 2--22). The algorithm first maps each update edge $\Delta e$ to a query edge and initializes the partial matches $M$ (lines 4--5). Subsequently, the DFS search starts from level {$l=2$} (line 6), as the first two query vertices are already determined during mapping. {The counter $p$ counts the candidate vertices being handled at layer $l$. Built on the current partial matches, WBM computes the candidate set for the next query vertex {using $\mathsf{GenCandidates}$}(line 7). $\mathsf{GenCandidates}$ initializes the candidates $res$ at layer $l$ to $CTable[\pi[l]]$ and filters it by intersecting with the neighbors of the data vertices matched.} The algorithm then proceeds to explore level by level (lines 8--22). When the search reaches the final level ($l = (|V(Q)|-1)$, i.e., all the vertices have been visited), the algorithm joins $M$ with each candidate vertex $c$ and adds to $\Delta M$ (lines 10--11). Then, WBM backtracks to the nearest preceding level with unexplored candidates (lines 12--13). Otherwise, it attempts to find a candidate vertex at the current level that can produce the candidate set for the next level (lines 15--20). If such a qualified candidate vertex is found (i.e., $C[l+1]\neq \O$), it appends ($\pi, C[l][p[l]]$) to $M$ and advances to the next level (lines 17--19). If there is no such candidate vertex, WBM backtracks to the previous level (lines 21--22). As each updated edge is handled by a warp, the threads responsible for the edge cooperatively compute the candidate vertices via $\mathsf{GenCandidates}$, i.e., executing the intersection and reading the global memory together. {Moreover, the intersection is implemented by parallel binary search and is efficient.} {It is worth noting that incremental matching can involve multiple updates, and due to the independent handling of each update, duplicates may arise. To eliminate duplicates, we assign a total order to each update and establish a rule during the matching process that dictates matches can only form from lower-order update edges to higher-order update edges~\cite{10.1145/3318464.3389699}.}

{
\textbf{Complexity.} For each update $\Delta e \in \Delta B$, procedure $\mathsf{GenCandidates}$ governs the execution time, which is implemented by binary search, leading to a time complexity $O(d_{max}(G)\log C_{max})$, where $d_{max}(G)$ and $C_{max}$ denotes the maximum degree of data vertex and the maximum count of candidate vertices, respectively. The intersection executes at most $d_max(Q)$ times, resulting in a worst case complexity $O_{gc} = O(d_max(Q)d_{max}(G)\log C_{max})$. Consequently, for each update batch $\Delta B$, the overall time complexity of Algorithm~\ref{alg:alg1} is $O(|\Delta B||E_Q|d_{max}(G)^{|V_Q-1|}d_max(Q)\log C_{max})$.
}

\begin{figure}[tb]
    \centering
    \includegraphics[width=0.30\textwidth]{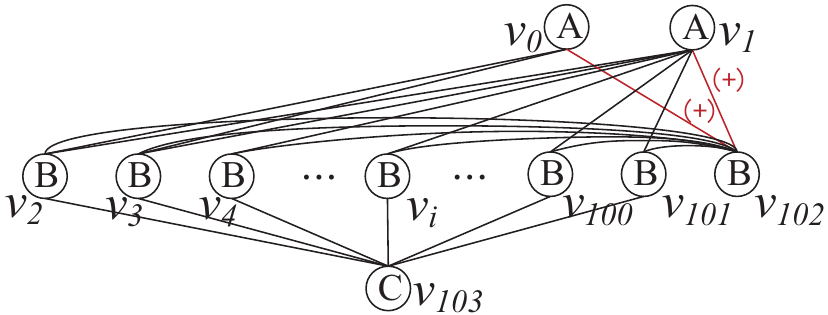} 
    \vspace{-3mm}
    \caption{{Updates with skewed workloads}}
    \label{fig:load_imbalance}
\end{figure}

\section{Practical Optimizations}
\label{sec:opt}

In this section, we delve into the intricacies of Algorithm~\ref{alg:alg1}, shedding light on its bottlenecks and presenting comprehensive insights into our strategic enhancements.

\subsection{Handling Load Imbalance}
\label{sec:handling_load_imbalance}

While the warp-centric assignment proves effective in alleviating thread and memory divergence, the persistent challenge lies in the {\em load imbalance} inherent to the DFS process. This stems from substantial {\em variations in the adjacency lists} of different vertices, making it impractical to precisely estimate the total number of neighbors for exploration. High-latency memory transactions are necessary to fetch multi-hop neighbors, adding to the complexity. Therefore, statically distributing tasks before execution becomes a formidable challenge. To tackle this, we propose a warp-centric work-stealing strategy aimed at resolving load imbalances among warps. Let's illustrate this issue further with an example.

\begin{example}
{
The data graph $G$ in {Figure~\ref{fig:load_imbalance}} contains two insertions, namely {blue}{$e(v_0,v_{102})$} and {blue}{$e(v_1,v_{102})$}, both matching the query edge $e(u_0,u_1)$ shown in {blue}{Figure~\ref{fig:load_imbalance}(a)}. {blue}{Figure~\ref{fig:example}(a)} illustrates the partial search tree corresponding to these insertions. 
The query vertices on the left side denote the matching order, where the first two vertices correspond to the query edge that matches the update edge. 
In this case, we initially map the insertions {blue}{$e(v_0,v_{102})$} and {blue}{$e(v_1,v_{102})$} to $e(u_0,u_1)$. In {blue}{Figure~\ref{fig:loadbalance}(a)}, we assign warp i and warp j to handle the updates {blue}{$e(v_0,v_{102})$} and {blue}{$e(v_1,v_{102})$}, respectively. 
As depicted in {blue}{Figure~\ref{fig:loadbalance}}, the number of qualified candidate vertices of $u_2$ and $u_3$ with respect to insertion {blue}{$(v_1,v_{102})$} is much higher than that of insertion {blue}{$(v_0,v_{102})$}. Consequently, {blue}{warp j} bears a heavier workload compared to  {blue}{warp i}, resulting in a notable workload imbalance. 
It's essential to note that, since the search is conducted in a DFS style rather than BFS, it's only possible to calculate the workloads after the search exhausts. Once {blue}{warp i} completes its assigned workloads, {blue}{warp j} still has unfinished workloads, indicating a significant imbalance.
}
\end{example}

As mentioned earlier, shared memory in a GPU is shared among threads within a block, offering fast accessibility and providing an opportunity to balance workloads among warps. We leverage the shared memory to implement our load balance optimization. There are two work stealing strategies to address the load imbalance issue.

\begin{itemize}
    \item \textbf{Passive Stealing.} We reserve an array within shared memory, whose length is equivalent to the count of warps present within a block. Each element in the array signifies whether a warp has completed its assigned workloads, initialized with zeros.
    As a warp concludes its workloads, it records its status by setting the corresponding element in the array to 1. 
    Periodically, warps with unfinished workloads scan the array to find an idle warp. When an idle warp is discovered, the warp with unfinished workloads undertakes a passive stealing operation, transferring a portion of its workloads to the idle warp. 
    Passive stealing allows the idle warps to acquiescently receive the appropriated workloads. However, this passive stealing can lead to thread underutilization because a warp must interrupt its ongoing workloads to search for an idle warp. 
    
    \item \textbf{Active Stealing.} To eliminate the need for periodic scrutiny of engaged warps, we propose the implementation of an active stealing strategy. 
    After a warp completes its current workloads, it inspects other warps within the same block to determine if any still have unfinished tasks, bypassing the need for a dedicated search for idle warps. 
    If such warps exist, the active stealing warp makes a thoughtful choice based on the remaining workloads and proceeds to actively appropriate half of its tasks. 
    This approach promotes a more efficient balance of workloads among warps within the block.
\end{itemize}

\begin{figure}[tb]
    \centering
    \includegraphics[width=0.48\textwidth]{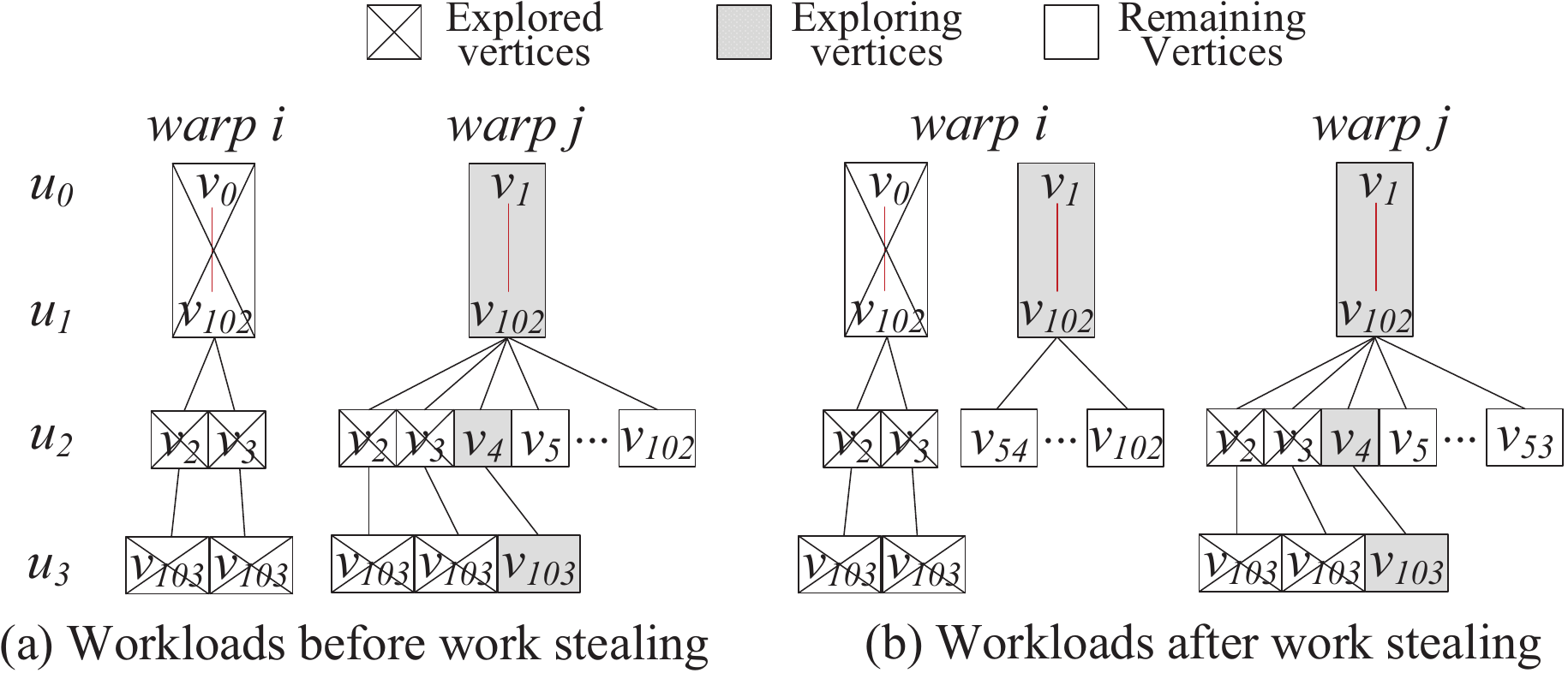}
    \vspace{-3mm}
    \caption{{Workloads of different warps and work stealing}}
    \label{fig:loadbalance}
\end{figure}

\begin{example}
{
In {Figure~\ref{fig:loadbalance}}, the active stealing strategy is visually presented. 
For simplicity, we assume that there are two warps in a block. Upon completing its workloads, a warp, let's say {blue}{warp i}, examines the variables $csize$ and $p$ stored in the shared memory layer by layer to detect those with unfinished workloads.
Within this process, {blue}{warp i iterates through the $csize$ and $p$ variables and successfully pinpoints unexplored candidates $\{v_5$,$\cdots$,$v_{102}\}$}. Subsequently, {blue}{warp i} appropriates half of the unexplored candidates, specifically {blue}{\{$v_{54}$,$\cdots$,$v_{102}$\}}, along with their parents {blue}{$v_1$} and {blue}{$v_{102}$}, and adds them to its own candidate set. 
Figure \ref{fig:loadbalance}(b) illustrates the resulting workloads of each warp after work stealing.
}
\end{example}

{
\textbf{Complexity.} Work stealing involves an idle warp sequentially scanning $csize$ and $p$, estimating remaining workloads layer by layer. This procedure takes $O(L|W|)$ time, where $L$ is the maximum number of layers to traverse and $|W|$ is the number of warps in a block.
}

{
Assuming workloads in a block are processed sequentially by a single warp with execution time of $T_{seq}$, work-stealing ideally ensures a fair workload distribution among warps, resulting in an execution time of $\frac{T_{seq}}{|W|}$ for each warp. This is equivalent to the total time due to the fair distribution.
Without work stealing, if the maximum execution time among warps in a block is $T_{max}$, the speedup can be expressed as $\frac{|W|T_{max}}{T_{seq}}$. Notably, work-stealing provides more significant speedup benefits with more skewed workloads (See \S~\ref{sec:scalability} and \S~\ref{sec:ablation} for experimental details).}
}

\subsection{Reducing Redundant Matching}
\label{sec:cs}

{When examining the incremental matches associated with the insertion $e(v_0,v_2)$ in Figure~\ref{fig:example}, it becomes apparent that the search traverses the same subgraphs induced by ${v_0,v_2,v_3}$ and ${ v_0,v_2,v_4 }$, resulting in redundant computations.} 
It is crucial to emphasize that both {these} induced data subgraphs and the induced query subgraph, comprising vertices $\{u_0,u_1,u_2\}$, exhibit automorphic properties, implying that they are isomorphic to themselves. 
For instance, a mapping $\{(u_0,u_0),(u_1,u_2),(u_2,u_1)\}$ can be found when considering the query subgraph induced by $\{u_0,u_1,u_2\}$. 
While previous studies have utilized automorphisms to reduce redundant computation in cases where the original query graph is automorphic, 
we can further minimize redundant computation by acknowledging the presence of automorphic subgraphs within the given query subgraph. Given a graph $g$, if $e(v_i,v_j), e(v_x,v_y) \in E(g)$, and $M(v_i)=v_x, M(v_j)=v_y$, then $e(v_i,v_j)$ and $ e(v_x,v_y)$ are equivalent. Hence, we define
\begin{definition}
\label{def:equivalent_edge_set}
An edge set $E^\prime(g) \subseteq E(g)$ is an equivalent edge set, if and only if any two edges $e(v_i,v_j)\in E^\prime(g)$ and $e(v_x,v_y) \in E^\prime(g)$ are equivalent.
\end{definition}

Given a graph $g$, if the induced subgraph obtained by removing $k$ vertices from $V(g)$ is automorphic, it is referred to as a \textbf{\textit{k}-degenerated automorphic subgraph} of $g$. The $k$-degenerated automorphic subgraph is denoted as $g^k=\{V^k$, $R^k$, $E^k$, $M^k\}$, where $V^k$ denotes the remaining vertices that constitute induced subgraph, which is automorphic. $R^k$ represents the removed vertex set, $E^k$ denotes the equivalent edge set w.r.t. the induced subgraph, and $M^k$ is the mapping of the induced subgraph.

\begin{example}
Consider the query graph $Q$ in Figure~\ref{fig:example}(a). After removing $u_3$, the induced subgraph comprised of $\{u_0,u_1,u_2\}$ is a $1$-degenerated automorphic subgraph of $Q$ and is denoted as $Q^1$, where $e(u_0,u_1)$ and $e(u_0,u_2)$ are equivalent. Specifically, $V^1=\{u_0,u_1,u_2\}$, $E^1=\{e(u_0,u_1)$, $e(u_0,u_2)\}$, $R^1=\{u_3\}$, and $M^1 = \{(u_0,u_0),(u_1,u_2),(u_2,u_1)\}$.
\end{example}

Thus, once a partial match $M$ is acquired for a query edge in the equivalent edge set $E^k$ of $Q^k$, we rearrange the query vertices in $M$ to generate other partial matches without traversing the same data subgraph multiple times. {For instance, given the partial match $M=\{(u_0,v_0),(u_1,v_2),(u_2,v_3)\}$ when mapping $e(v_0,v_2)$ to $e(u_0,u_1)$, since $e(u_0,u_1)$, $e(u_0,u_2)$ $\in E^1$ of $Q^1$, according to $M^1$, we interchange the positions of $u_1$ and $u_2$ in $M$ to obtain another partial match $M^\prime = \{(u_0,v_0)$, $(u_2,v_2)$, $(u_1,v_3)\}$}. 
Consequently, the search for the same data subgraph is conducted only once. This permutation operation is referred to {\textbf{coalesced search}, as it coalesces the search for the partial matches of multiple permutations of $V^k$ in $Q^k$.}  

To generate the set $\{g^k\}$, we begin with $k=0$ and progressively remove $k$ vertices to generate the induced subgraphs with different sizes. Subsequently, we examine whether each induced subgraph is automorphic. However, as a query edge $e$ can belong to multiple entries in $\{g^k\}$, redundant matching occurs. Therefore, if an edge $e$ is found in both $g^{k_i}_i, g^{k_j}_j\in \{g_k\}$, we heuristically apply the following rules:
\begin{enumerate}
    \item If $k_i < k_j$, exclude $e$ from $E^{k_j}_j$ of $g^{k_j}_j$;
    \item If $k_i = k_j$, select $g^{k_i}_i$ if $|E^{k_i}_i| > |E^{k_j}_j|$.
\end{enumerate}

Rule 1 aims to ensure that a larger data subgraph can be shared, reducing redundant matching. By excluding the common edge $e$ from $E^{k_j}_j$ when $k_i < k_j$, we allow for the inclusion of a larger data subgraph during the matching process. Rule 2, on the other hand, focuses on maximizing the number of edges that can be shared. By selecting $g^{k_i}_i$ with a larger $|E^{k_i}_i|$ size when $k_i = k_j$, we prioritize the inclusion of more edges during the matching, leading to a more efficient exploration. Both rules contribute to reducing redundancy and improving the efficiency of matching by promoting the sharing of larger data subgraphs and maximizing the sharing of edges.

\textbf{Avoid Invalid Matching.} Considering $Q^1$ of $Q$ in Figure~\ref{fig:example}, if we first map an updated edge to $e(u_0,u_1)$, we can obtain another valid partial match by interchanging the positions of $u_1$ and $u_2$. However, an invalid partial match can be generated if we first map the updated edge to $e(u_0,u_2)$, since a data vertex that matches $u_1$ should also possess neighbors labeled as C (which is not the case for $u_2$). In such a situation, we say $e(u_0,u_1)$ dominates $e(u_0,u_2)$ and designate $e(u_0,u_1)$ as a prioritized query edge. For each $Q^k$, we prioritize the matching of the prioritized query edge to avoid invalid matches.

{
\textbf{Remark.} Assuming that an update $\Delta e \in \Delta B$ maps to an edge $e^k \in E^k$, discovering an arbitrary partial match for $V^k$ immediately yields $|E^k-1|$ additional partial matches for the remaining edges in $E^k$ through permutation. This leads to a potential speedup of $|E^k|$. 
However, query vertices in $V^k$ adjacent to removed vertices may lose specific label constraints, expanding the candidate space for the remaining vertices. To address this, we selectively eliminate isolated query vertices with a degree of 1. 
Consolidating partial matches involving these vertices through parallel join operations proves significantly more efficient than individual vertex mappings. While the impact is minimal compared to DFS-based exploration, this practical approach ensures a speedup within the range of $(1, |E^k|)$ w.r.t the time complexity of Algorithm~\ref{alg:alg1}.
}

\subsection{Practical Implementation and Optimizations}

\label{sec:data_structure}
We further explore the integration of our system with existing dynamic graph data structures and how we address any inherent flaws that may arise.

\textbf{Graph Container.} 
\label{sec:graph_container}
{
In our study, we adopt GPMA~\cite{sha2017accelerating} as the foundational data structure for its simplicity and efficiency. GPMA employs a sorted array, PMA, to manage edges by reserving spaces based on upper and lower thresholds. Each edge in a batch of updates is assigned a thread, locating the leaf segments to which it belongs. 
GPMA then groups edges within the same segment and materializes updates if the segment has sufficient space within the thresholds. If not, GPMA resorts to its parent, composed of two adjacent segments, in a bottom-up iterative process until all updates are processed. 
}

{
\textbf{Other Practical Optimizations.} 
As the data structure resides in global memory and involves multiple threads executing memory access during the location step, minimizing the associated overhead is crucial. To address this, we optimize by loading the top-$k$ layers into shared memory for efficient reading. 
Additionally, GPMA employs tailored strategies for insertions based on segment sizes. It utilizes the warp technique for segments up to 32 elements, employs the block method for segments fitting into shared memory, and resorts to the device approach for larger segments exceeding shared memory capacity.
While the warp-based optimization may lead to suboptimal parallelism for segments with fewer than 32 elements, we leverage the Cooperative Group (CG)~\cite{cooperative_group} to enhance flexibility in thread grouping within a warp. By partitioning the warp into smaller thread groups based on powers of 2 (e.g., 16, 8, etc.) and allocating them based on segment sizes, we address this issue. For instance, for segments in the 16 to 32 range, we initiate processing with 16-sized thread groups and adaptively allocate smaller groups as entries are processed. This adaptive strategy improves thread utilization and mitigates the stall issue.
}

\section{Experiments}
\label{sec:experiments}

In this section, we evaluate the performance of the proposed system and conduct a comparative evaluation with existing state-of-the-art dynamic subgraph matching methods.
\subsection{Experimental Setup}

\noindent \textbf{Datasets}. {We employ six datasets~\cite{10.14778/3551793.3551803,yang2023fast} in our experiments.} Detailed dataset descriptions are provided in Table~\ref{tab:datasets}, where $|V|$ and $|E|$ refer to the numbers of vertices and edges, respectively. $\Sigma_V$ and $\Sigma_E$ represent the quantities of vertex labels and edge labels, respectively. $d_{avg}$ is the average. 
{We set the insertion/deletion rate, namely the batch size, in the range of 2\% to 10\%, with \underline{\textbf{10\%}} being the default value.} The insertion/deletion rate represents the proportion of the number of inserted/deleted edges to the total number of edges in the data graph.

\begin{table}[t]
    \tiny
    \centering
    \caption{Summary of the datasets } 
    \vspace{-3mm}
    \label{tab:datasets}    
    \resizebox{\linewidth}{!}{
    \begin{tabular}{|c|c|c|c|c|c|}
    \hline
    \textbf{Datasets} & $\boldsymbol{|V|}$ & $\boldsymbol{|E|}$ & $\boldsymbol{|\Sigma_V|}$ & $\boldsymbol{|\Sigma_E|}$ & $\boldsymbol{d_{avg}}$  \\ \hline\hline
    {Github (\textit{GH})}      & {37.7K} & {0.3M}  &{5}   &{1}   &{15.3}      \\ \hline
    {Skitter (\textit{ST})}      & {1.7M} & {11.1M}  &{25}   &{1}   &{13.1}      \\ \hline
    Amazon (\textit{AZ})      & 0.4M & 2.4M  & 6  & 1  & 12.2     \\ \hline
    LiveJournal (\textit{LJ}) & 4.9M & 42.9M & 30 & 1  & 18.1    \\ \hline
    Netflow (\textit{NF})     & 3.1M & 2.9M  & 1  & 7  & 2.0       \\ \hline
    LSBench (\textit{LS})     & 5.2M & 20.3M & 1  & 44 & 8.2      \\ \hline
    \end{tabular}}
 \vspace{-6mm}
\end{table}

\noindent \textbf{Queries}. Following precedent studies~\cite{Sun:2020:ISC:3399666.3399897,10.14778/3551793.3551803}, we generate query graphs by randomly extracting subgraphs from the data graph. {The query graphs are categorized into {Dense} ($d_{avg} \geq 3$), {Sparse} ($d_{avg} < 3$), and {Tree} ($d_{avg}=|V_Q|-1$), with $d_{avg}$ representing the average degree of queries.} For each type, we create query graphs while varying the number of vertices, $|V_Q|$, from 4 to 12. We produce a query set consisting of 50 query graphs of each size and type. By default, we present the results for query sets composed of query graphs with \underline{\textbf{6}} vertices.

\noindent \textbf{Baseline Methods}. We conduct a comparative analysis between our {method} and state-of-the-art continuous subgraph graph matching algorithms, which encompass {TurboFlux (\textsf{TF})~\cite{10.1145/3183713.3196917}, SymBi (\textsf{SYM})~\cite{10.14778/3457390.3457395}, RapidFlow (\textsf{RF})~\cite{10.14778/3551793.3551803} and CaLig (\textsf{CL})~\cite{yang2023fast}}. Notably, as far as our knowledge extends, there exists no algorithm tailored for batch-dynamic graph subgraph matching specifically designed for the GPU architecture.

\noindent \textbf{Running Platform}. The experimental evaluations are carried out on an Ubuntu server, which is equipped with an Intel Core i9-10900X CPU and 128GB of host memory. Additionally, the server features an Nvidia Geforce RTX 3090 GPU, boasting 24GB of device memory and 83 streaming multiprocessors.

\noindent \textbf{Metrics}. The performance evaluation of each algorithm is based on the average query latency across all query graphs. To avoid excessively long running times, a time threshold of 30 minutes is set. If the execution of a query surpasses this threshold, it is terminated and classified as an unsolved query. Consequently, the percentage of solved queries is also reported as a performance metric. The average query time excludes queries that exceed the time threshold. In addition, we use GPU utilization to evaluate the performance of our method.

\begin{table}[t]
    \centering
    \caption{Overall performance compared with baselines}
    \vspace{-3mm}
    \label{tab:overallperformance}
    \resizebox{\columnwidth}{!}{
    \renewcommand{\arraystretch}{1.3}
    
    \begin{tabular}{|c|c|clclclclcl|}
    \hline
    \multirow{2}{*}{\textbf{QS}} & \multirow{2}{*}{\textbf{DS}} & \multicolumn{10}{c|}{\textbf{Method}}                                                                                                                                            \\ \cline{3-12} 
                                 &                              & \multicolumn{2}{c|}{\textsf{TF}} & \multicolumn{2}{c|}{{\textsf{SYM}}} & \multicolumn{2}{c|}{{\textsf{RF}}} & \multicolumn{2}{c|}{{\textsf{CL}}} & \multicolumn{2}{c|}{{\textsf{\textsf{GAMMA}}}} \\ \hline\hline
    \multirow{6}{*}{Dense}       & \textit{GH}                           & \multicolumn{2}{c|}{127.746(5)\tnote{a}}      & \multicolumn{2}{c|}{3.755}        & \multicolumn{2}{c|}{0.202}       & \multicolumn{2}{c|}{35.082}      & \multicolumn{2}{c|}{0.553}          \\ \cline{2-12}
                                 & \textit{ST}                           & \multicolumn{2}{c|}{86.813}            & \multicolumn{2}{c|}{10.298}       & \multicolumn{2}{c|}{0.228}       & \multicolumn{2}{c|}{48.933 (1)}   & \multicolumn{2}{c|}{0.355}               \\ \cline{2-12} 
                                 & \textit{AZ}                           & \multicolumn{2}{c|}{7.251}       & \multicolumn{2}{c|}{1.385}        & \multicolumn{2}{c|}{0.328}       & \multicolumn{2}{c|}{7.291}       & \multicolumn{2}{c|}{0.469}          \\ \cline{2-12} 
                                 & \textit{LJ}                           & \multicolumn{2}{c|}{13.678}      & \multicolumn{2}{c|}{15.666}       & \multicolumn{2}{c|}{0.757}       & \multicolumn{2}{c|}{20.372}      & \multicolumn{2}{c|}{0.611}          \\ \cline{2-12} 
                                 & \textit{NF}                           & \multicolumn{2}{c|}{3.557 (1)}    & \multicolumn{2}{c|}{2.068}        & \multicolumn{2}{c|}{0.185}       & \multicolumn{2}{c|}{0.711(16)}       & \multicolumn{2}{c|}{0.51}          \\ \cline{2-12} 
                                 & \textit{LS}                           & \multicolumn{2}{c|}{3.216}       & \multicolumn{2}{c|}{3.979}        & \multicolumn{2}{c|}{0.443}       & \multicolumn{2}{c|}{73.205(20)}       & \multicolumn{2}{c|}{0.473}          \\ \hline\hline
    \multirow{6}{*}{Sparse}      & \textit{GH}                           & \multicolumn{2}{c|}{662.613(29)
}     & \multicolumn{2}{c|}{493.747 (12)}  & \multicolumn{2}{c|}{140.979}     & \multicolumn{2}{c|}{401.436 (12)} & \multicolumn{2}{c|}{8.110}          \\ \cline{2-12}
                                & \textit{ST}                           & \multicolumn{2}{c|}{269.747(23)}            & \multicolumn{2}{c|}{257.793(16)}       & \multicolumn{2}{c|}{66.325}       & \multicolumn{2}{c|}{139.643 (4)
}   & \multicolumn{2}{c|}{7.218}               \\ \cline{2-12} 
                                 & \textit{AZ}                           & \multicolumn{2}{c|}{10.548}      & \multicolumn{2}{c|}{2.155}        & \multicolumn{2}{c|}{0.325}       & \multicolumn{2}{c|}{2.431}       & \multicolumn{2}{c|}{0.669}          \\ \cline{2-12} 
                                 & \textit{LJ}                           & \multicolumn{2}{c|}{39.857}      & \multicolumn{2}{c|}{14.076}       & \multicolumn{2}{c|}{0.680}       & \multicolumn{2}{c|}{1.165}       & \multicolumn{2}{c|}{0.715}          \\ \cline{2-12} 
                                 & \textit{NF}                           & \multicolumn{2}{c|}{238.746 (13)} & \multicolumn{2}{c|}{65.609 (6)}    & \multicolumn{2}{c|}{1.612}       & \multicolumn{2}{c|}{1800(50)}       & \multicolumn{2}{c|}{0.99}          \\ \cline{2-12} 
                                 & \textit{LS}                           & \multicolumn{2}{c|}{63.706 (17)}  & \multicolumn{2}{c|}{34.658}       & \multicolumn{2}{c|}{4.730}       & \multicolumn{2}{c|}{1800(50)}      & \multicolumn{2}{c|}{1.469}          \\ \hline\hline
    \multirow{6}{*}{Tree}        & \textit{GH}                           & \multicolumn{2}{c|}{745.289(41)
}     & \multicolumn{2}{c|}{ 1225.02(45)}      & \multicolumn{2}{c|}{366.981}     & \multicolumn{2}{c|}{744.025(22)}     & \multicolumn{2}{c|}{ 23.647}         \\ \cline{2-12}
                                 & \textit{ST}                           & \multicolumn{2}{c|}{1543.516(46)}            & \multicolumn{2}{c|}{ 1107.324(41)}       & \multicolumn{2}{c|}{498.218}       & \multicolumn{2}{c|}{ 382.392(36)}   & \multicolumn{2}{c|}{33.465}               \\ \cline{2-12} 
                                 & \textit{AZ}                           & \multicolumn{2}{c|}{117.9(1)}  & \multicolumn{2}{c|}{245.124(2)}   & \multicolumn{2}{c|}{ 7.767}       & \multicolumn{2}{c|}{ 122.538}     & \multicolumn{2}{c|}{9.791}          \\ \cline{2-12} 
                                 & \textit{LJ}                           & \multicolumn{2}{c|}{97.283 (1)}   & \multicolumn{2}{c|}{231.416 (1)}   & \multicolumn{2}{c|}{8.473}       & \multicolumn{2}{c|}{67.651 (3)}   & \multicolumn{2}{c|}{2.049}          \\ \cline{2-12} 
                                 & \textit{NF}                           & \multicolumn{2}{c|}{323.534(140)} & \multicolumn{2}{c|}{452.288 (15)}  & \multicolumn{2}{c|}{42.850(8)}   & \multicolumn{2}{c|}{ 1800(50)} & \multicolumn{2}{c|}{5.369 (2)}       \\ \cline{2-12} 
                                 & \textit{LS}                           & \multicolumn{2}{c|}{119.361 (7)}  & \multicolumn{2}{c|}{140.030 (7)}   & \multicolumn{2}{c|}{15.434 (6)}   & \multicolumn{2}{c|}{ 1800(50)}   & \multicolumn{2}{c|}{5.384 (3)}       \\ \hline
    
    \end{tabular}}
    \begin{tablenotes}
    \item \hspace{-3mm}*Values outside parentheses represent the average query latency (s), while those within indicate the number of unsolved queries.
    \end{tablenotes}
  \vspace{-8mm}
  
\end{table}

\subsection{Overall Performance}
Table~\ref{tab:overallperformance} summarizes the performance comparisons. In this context, queries exceeding the predefined time limit have been excluded. The values outside parentheses signify the average query latency in seconds, while the values within parentheses indicate the count of unsolved queries. QS and DS respectively denote the dataset and query structure.

{
The first observation is that both the average query latency and the quantity of unresolved queries typically increase as query density decreases. This is attributable to the prevalence of power-law distributions in real-world graphs, wherein vertices with low degrees predominate, thus culminating in a notably substantial count of query results on sparse queries. Furthermore, as the query graph becomes sparser, it proffers fewer pruning constraints, specifically interconnections among vertices, resulting in a considerably expanded search space. Second, although \textit{LJ} and \textit{LS} are of comparable size, the performance on these two datasets diverges markedly. \textit{LJ} presents a greater challenge when dealing with dense queries, while \textit{LS} proves more formidable in handling sparse and tree queries. The underlying rationale is that \textit{LJ} boasts a substantially higher average degree, thereby yielding a profusion of matches on dense queries, whereas \textit{LS}, conversely, possesses a lower average degree, culminating in an abundance of matches on sparse and tree queries. Third, most of the methods exhibit longer query latency on \textit{NF} compared to \textit{LS}, despite \textit{NF} having a smaller size. One possible reason is the highly skewed edge labels in \textit{NF}, which in turn enlarges the search space. \textsf{CL} exhibits poor performance on edge-labeled graphs, primarily due to its requirement to transform them into vertex-labeled graphs. The transformation treats labeled edges as labeled vertices connecting two endpoints, thereby altering the graph structure and expanding the search space. \textsf{RF} outperforms both existing methods on all query sets, owing to the proposed query reduction and dual matching techniques. These techniques effectively eliminate invalid partial results and redundant computations caused by automorphisms in query graphs.} Fourth, leveraging the abundant parallel potential offered by GPU and incorporating various practical optimization techniques, \textsf{\textsf{GAMMA}} demonstrates competitive or even optimal performance compared to existing algorithms across most query sets. The observed speedup ratio ranges from several times to tens of times when compared to the best baseline. In particular, our proposed method showcases substantial performance improvements of 67$\times$, 33$\times$, 5$\times$ and 712$\times$ on average, compared to \textsf{TF}, \textsf{SYM}, \textsf{RF} and \textsf{CL}, respectively. However, for some shorter-running queries, \textsf{\textsf{GAMMA}} demonstrates comparable performance with that of \textsf{RF}. This phenomenon arises from the limitation of such queries to sufficiently saturate the GPU with tasks. Moreover, \textsf{\textsf{GAMMA}} significantly reduces the number of unsolved queries compared to the baselines, with the majority of unsolved queries being tree queries resulting from the abundance of matches. This outcome serves as a clear indication of the scalability of our proposed method.

\begin{figure*}[tb]
    \centering
    \includegraphics[width=0.70\textwidth]{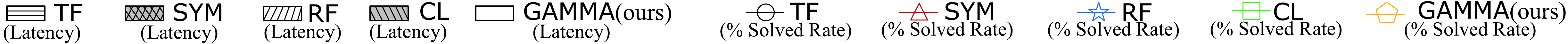}
    \vspace{1mm}
    \hspace{-5mm}
     \subfigure[Queries on \textit{GH}]{
    \includegraphics[width=0.5\textwidth]{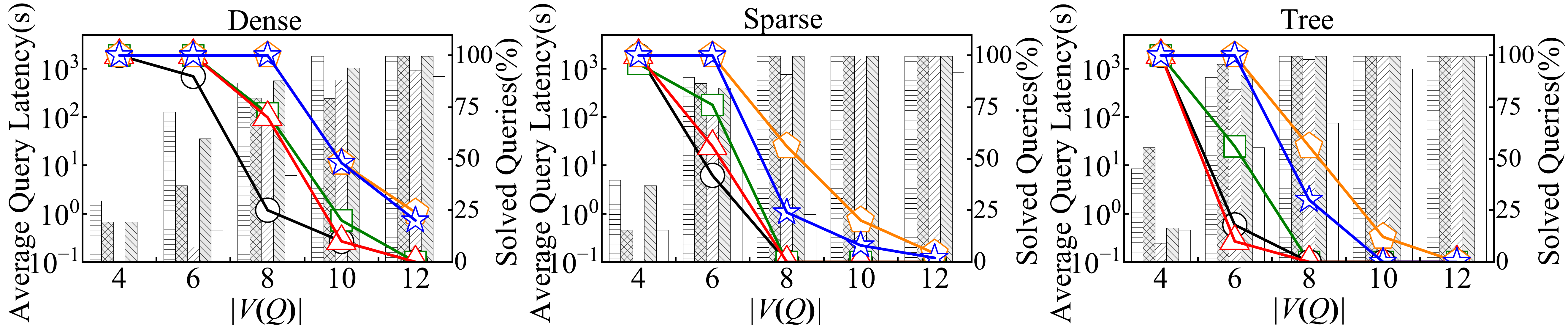}}
    \hspace{-2mm}
    \subfigure[Queries on \textit{ST}]{
    \includegraphics[width=0.5\textwidth]{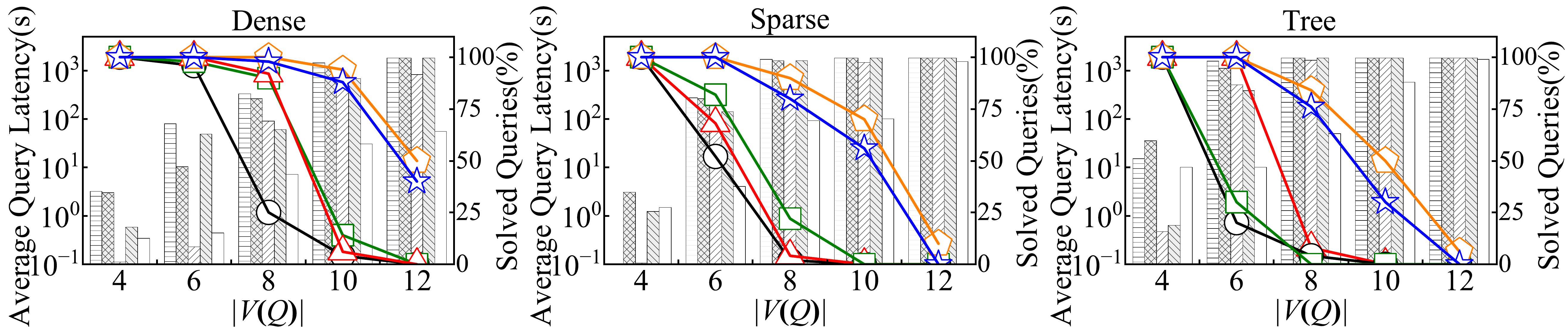}
    }
    
    \vspace{-4mm}
    \caption{{Scalability evaluation vs. query graph size $|V(Q)|$}}
    \label{fig:vary_size}
    \vspace{-2mm}
\end{figure*}
 
\begin{figure*}[tb]
    \centering
    \includegraphics[width=0.70\textwidth]{legend_overall.pdf}
    \vspace{1mm}
    \hspace{-5mm}
     \subfigure[Queries on \textit{GH}]{
    \includegraphics[width=0.5\textwidth]{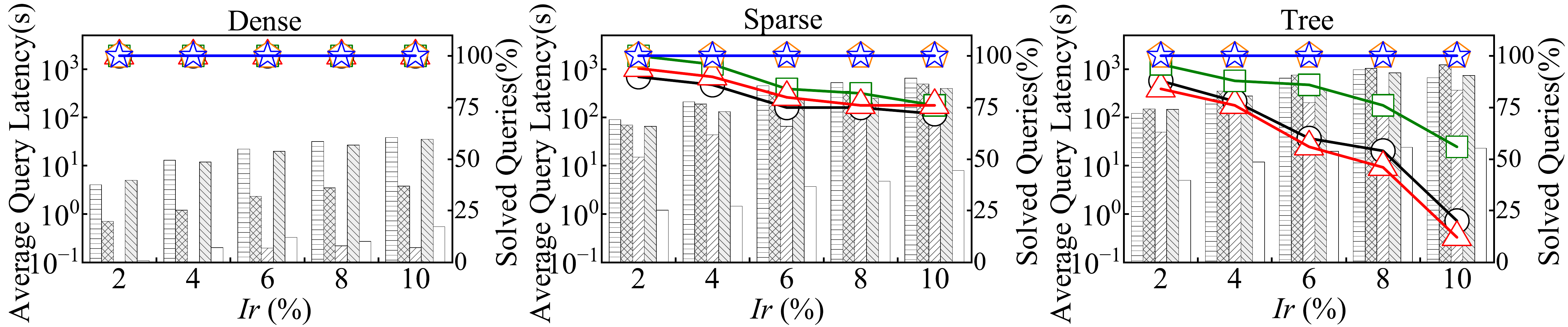}}
    \hspace{-2mm}
    \subfigure[Queries on \textit{ST}]{
    \includegraphics[width=0.5\textwidth]{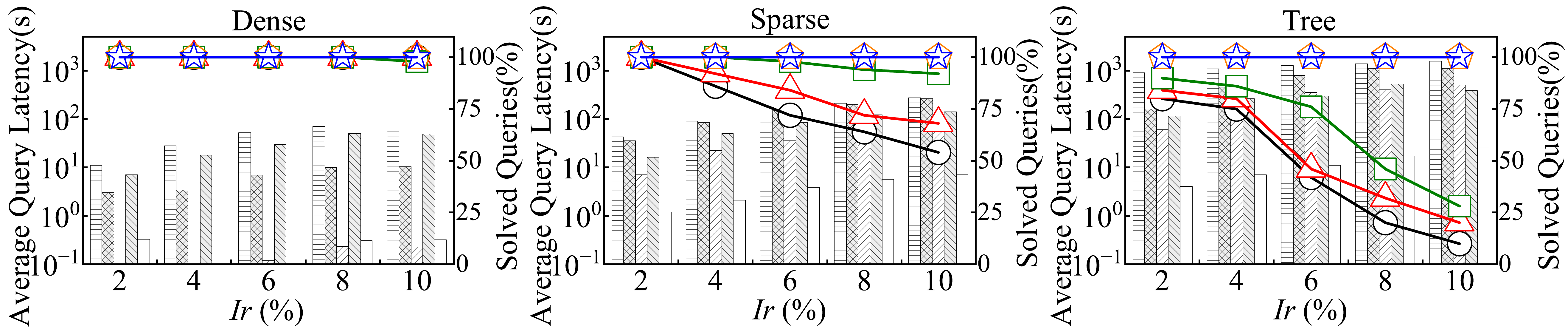}
    }

    \vspace{-4mm}
    \caption{{Scalability evaluation vs. insertion rate $I_r$}}
    \label{fig:vary_rate}
\vspace{-5mm}
\end{figure*}




\subsection{Scalability}
\label{sec:scalability}
We proceed to assess the impact of query graph size and insertion rate on performance. {Due to space constraints and similar experimental outcomes, we only report the experimental results on \textit{GH} and \textit{ST}. The version with complete experimental results on all datasets is available at https://github.com/ZJU-
DAILY/GAMMA/blob/main/GAMMA.pdf.}

\begin{figure}[tb]
    \centering
    \includegraphics[width=0.335\textwidth]{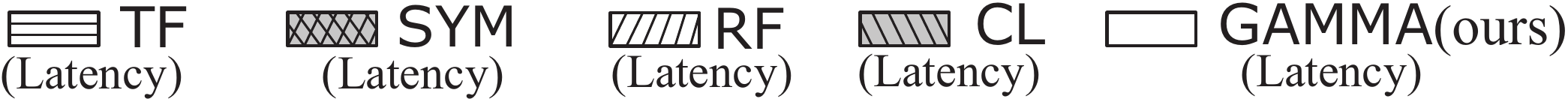} 
    
    \includegraphics[width=0.49\textwidth,height=0.12\textwidth]{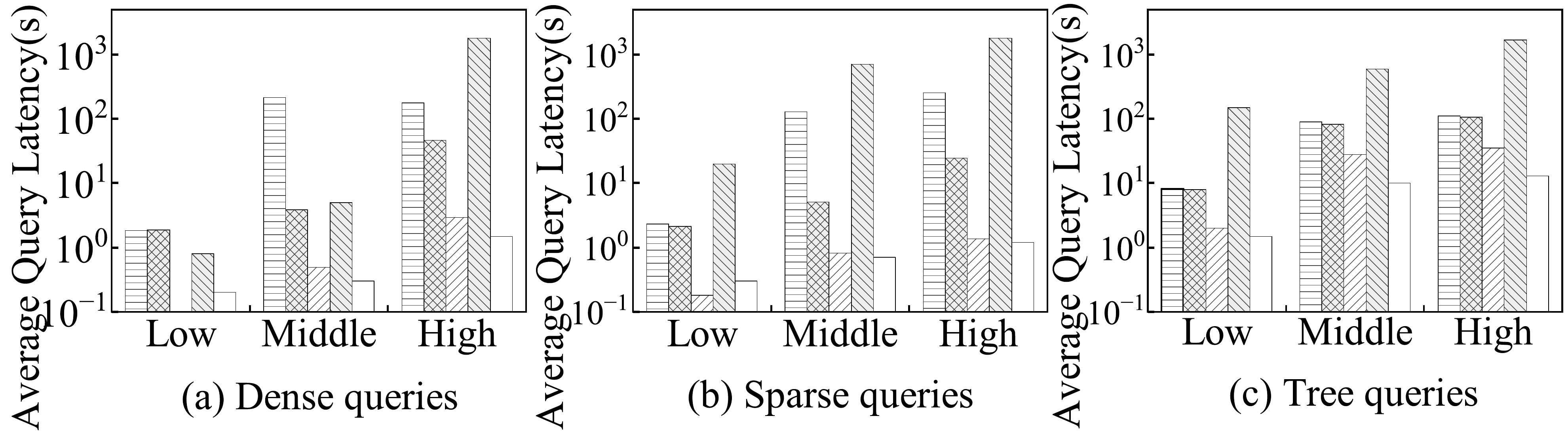} 
    
    \vspace{-3mm}
    \caption{{Scalability evaluation vs. density on \textit{LS}}}
    \label{fig:density}
\vspace{-6mm}
\end{figure}

\textbf{{Varying Query Graph Size.}} We assess the performance by varying the query graph size $|V(Q)|$ from 4 to 12. Figure~\ref{fig:vary_size} illustrates the corresponding results, where \textsf{\textsf{GAMMA}} consistently achieves the best performance. As observed, the average query latency generally increases, and the number of unsolved queries rises with the expansion of the query size due to the larger exploration space. With increasing query size, the performance gap between the baselines and our proposed method progressively widens. This phenomenon occurs because \textsf{\textsf{GAMMA}} effectively explores the large search space in parallel, while the baselines conduct the search sequentially. As the exploration space expands with the growth of graph size, the parallel approach becomes more crucial in showcasing the superiority of \textsf{\textsf{GAMMA}}. Moreover, \textsf{\textsf{GAMMA}} significantly outperforms the baselines in terms of the number of solved queries, especially concerning larger queries.

\textbf{{Varying Insertion Rate.}} We explore the performance under different insertion rates, ranging from 2\% to 10\%. The experimental results are presented in Figures \ref{fig:vary_rate}. We have excluded the experimental results for varying deletion rates, as they exhibit a similar pattern. In general, the query time increases as the insertion rate rises. This is particularly evident when updating the indices of the baselines, as they involve the edges of the data graph. Consequently, the query time for a single update grows as the insertion progresses. On the other hand, RF updates the indices by taking into account both the query graph size and the average degree of the data graph. It conducts matching by leveraging the local index with a better matching order. By efficiently utilizing the parallelism offered by the GPU to amortize the query overhead, our proposed method achieves improved scalability.

{
\textbf{Varying Density.} We evaluate how the density of the update regions impacts the performance. Following the previous study~\cite{10.14778/3551793.3551803}, we perform $k$-core decomposition on \textit{LS} and sample edges from these cores for insertions. We vary {$k\in \{4,8,12\}$} to depict the density of regions (i.e., low, middle, and high density). Figure~\ref{fig:density} illustrates the results. As observed, the runtime of all methods increases with the growth of density. Notably, our system exhibits a more pronounced acceleration in denser regions, courtesy of heightened parallelism and an optimally distributed workload.}




\begin{figure}
    \centering
    \includegraphics[width=0.335\textwidth]{legend_density.pdf} 
  \begin{minipage}[t]{0.6\linewidth}
    \centering
    
    \includegraphics[scale=0.115]{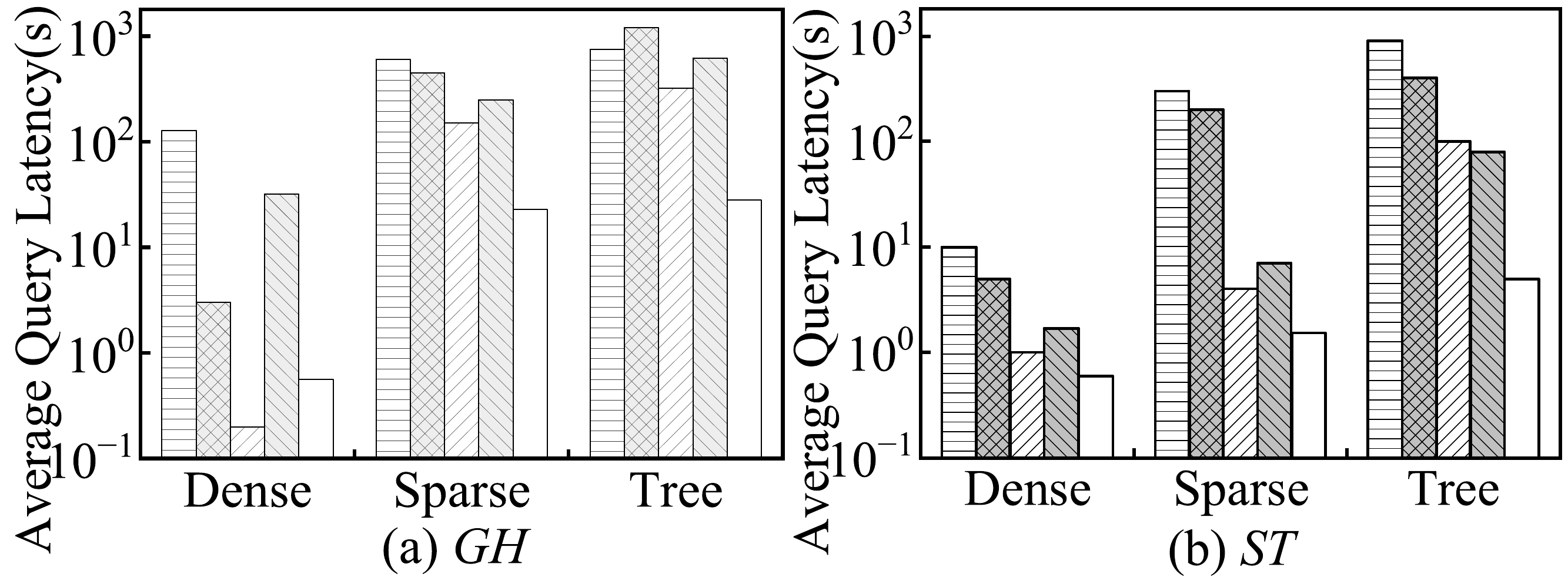}
    \vspace{-4mm}
    \caption{{Performance vs. mixed workloads}}
    \label{fig:mix}
  \end{minipage}%
  \begin{minipage}[t]{0.4\linewidth}
    \centering
    \includegraphics[scale=0.11]{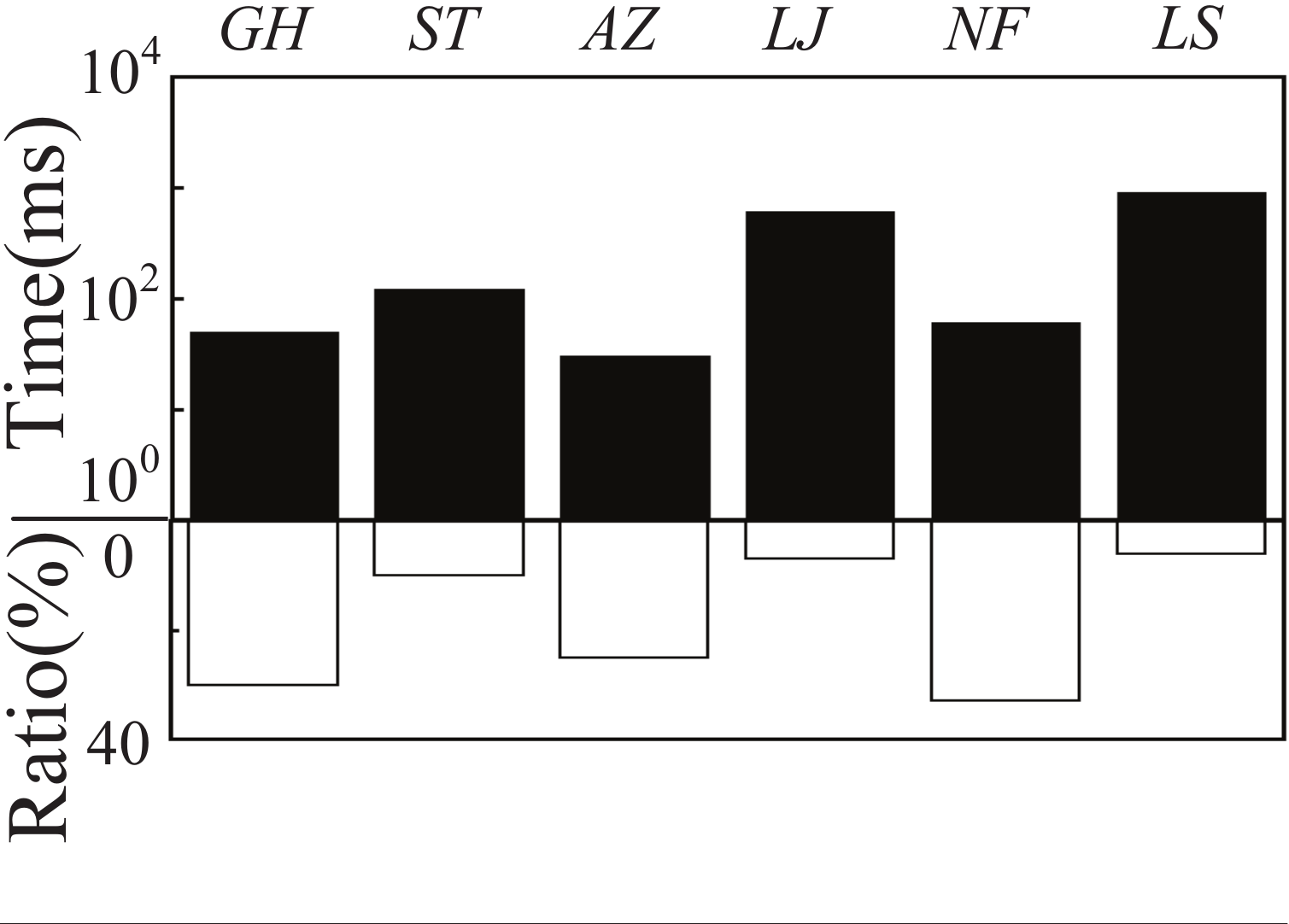}
    \vspace{-4mm}
    \caption{{Preprocessing analysis}}
    \label{fig:preprocess}
  \end{minipage}
  \vspace{-8mm}
\end{figure}

 {
\textbf{Mixed Workloads.} We evaluate the performance under mixed workloads of insertions and deletions. Following~\cite{yang2023fast}, we set the insertion-to-deletion ratio to $2:1$. The experimental results are presented in Figure~\ref{fig:mix}. Similar to the single workload scenario (i.e., Figure~\ref{fig:vary_rate}), the runtime of all the methods increases as the query set density decreases and our method outperforms all the competitors, showcasing the scalability of our approach under mixed workloads.}

{\textbf{Efficiency of Preprocessing.} We evaluate the performance of preprocessing.  Preprocessing consists of CPU-based candidate generation and GPU-based graph update. The candidate generation operates asynchronously with GPU execution and proves to be efficient. Consequently, the key factor affecting overall preprocessing performance is the graph update, which runs alongside incremental matching. Figure~\ref{fig:preprocess} provides insights into the graph update {at a 10\% update rate}, where `Time' denotes the graph update time and `Ratio' represents the proportion of graph update time relative to the total running time. Notably, the graph update is primarily impacted by the volume of updates. As depicted, a larger data size, thereby a larger volume of updates, results in more pre-processing time. 
}
 
\begin{figure*}[tb]
    \centering
    \includegraphics[width=0.75\textwidth]{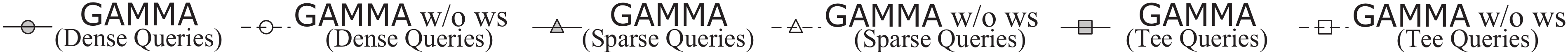}\\

    \hspace{-2mm}
    \subfigure[\textit{GH}]{
    \includegraphics[width=0.24\textwidth]{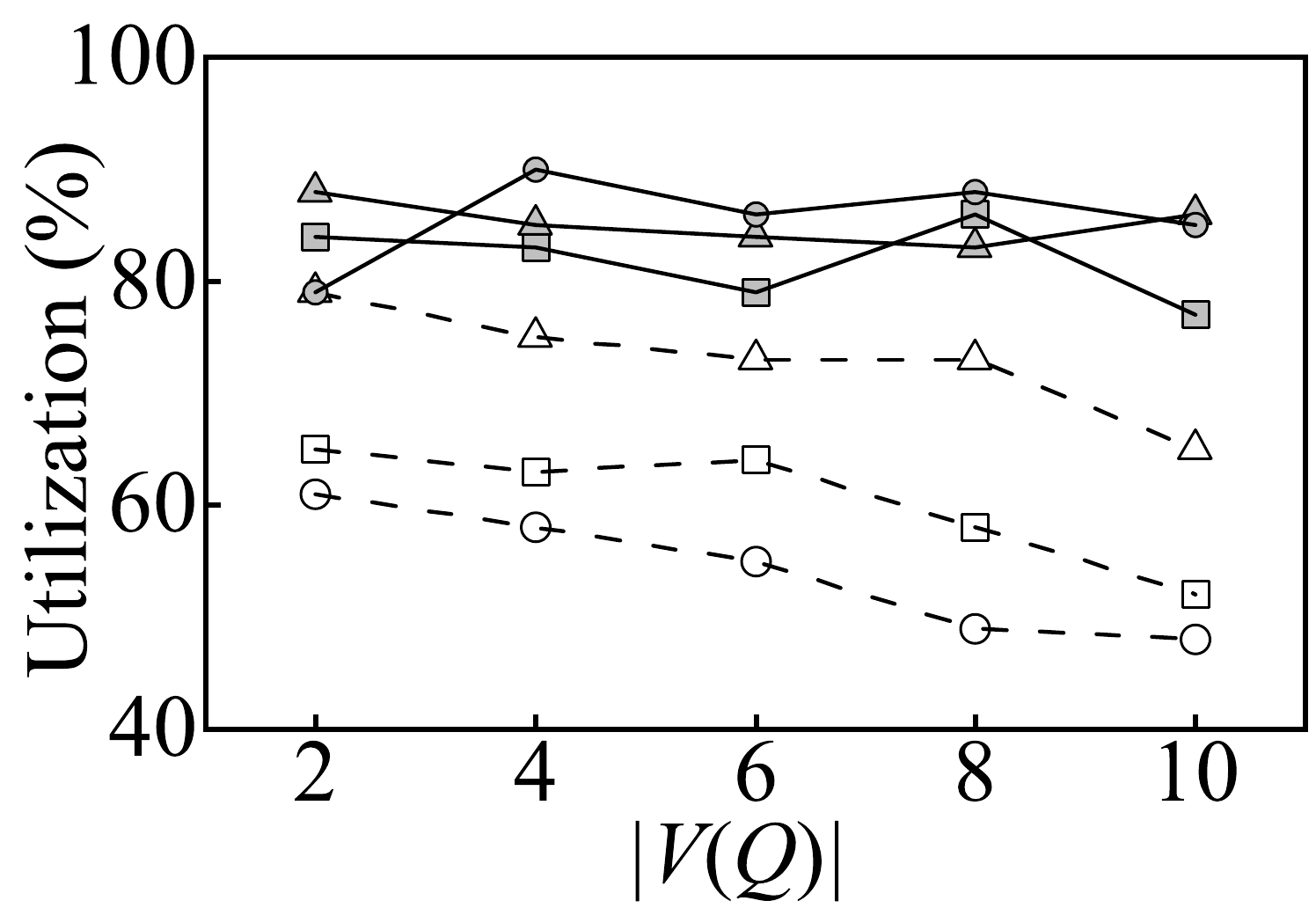}}
    \hspace{-3mm}
    \subfigure[\textit{ST}]{
    \includegraphics[width=0.24\textwidth]{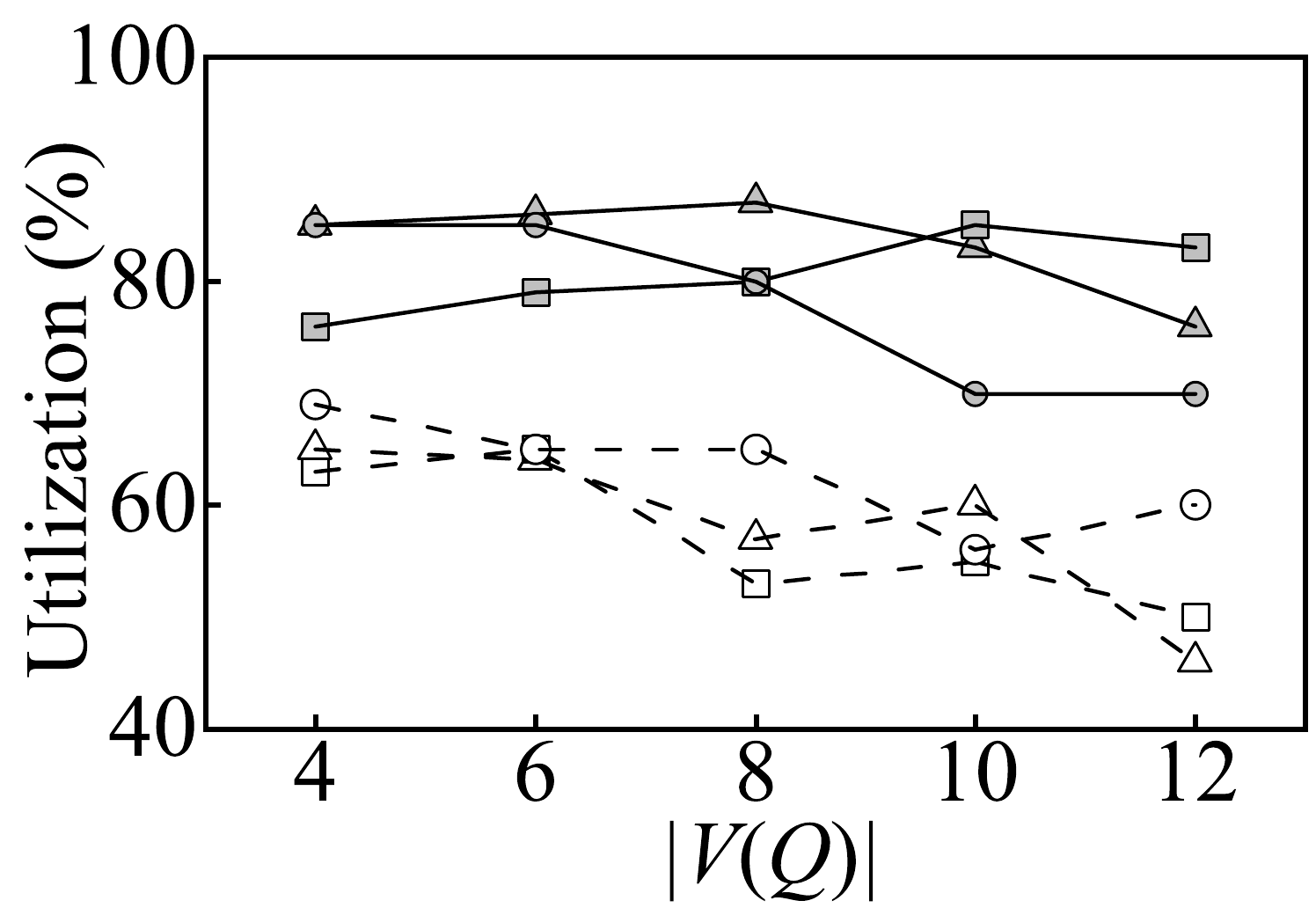}
    }
    \hspace{-3mm}
    \subfigure[\textit{GH}]{
    \includegraphics[width=0.24\textwidth]{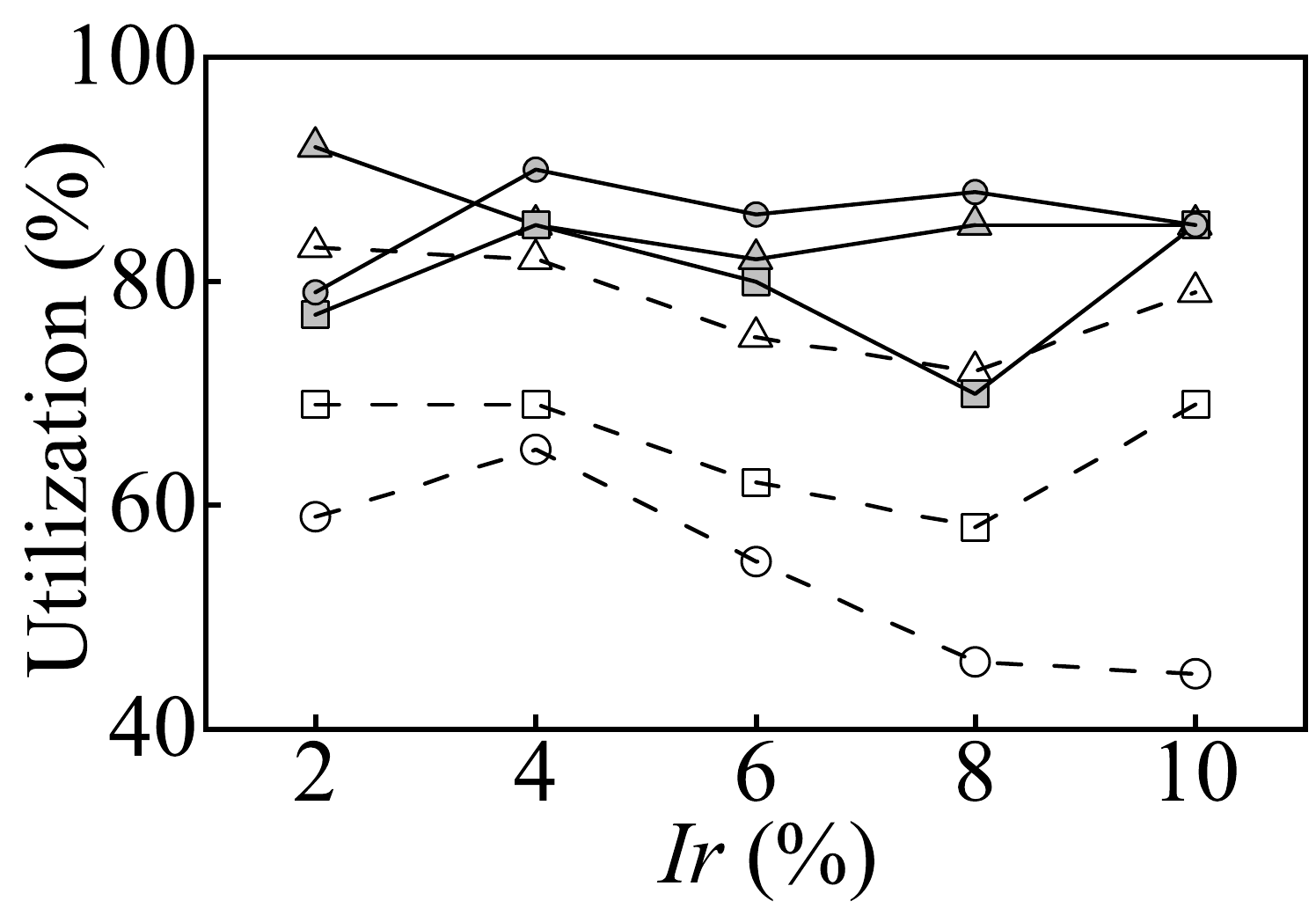}
    }
    \hspace{-3mm}
    \subfigure[\textit{ST}]{
    \includegraphics[width=0.24\textwidth]{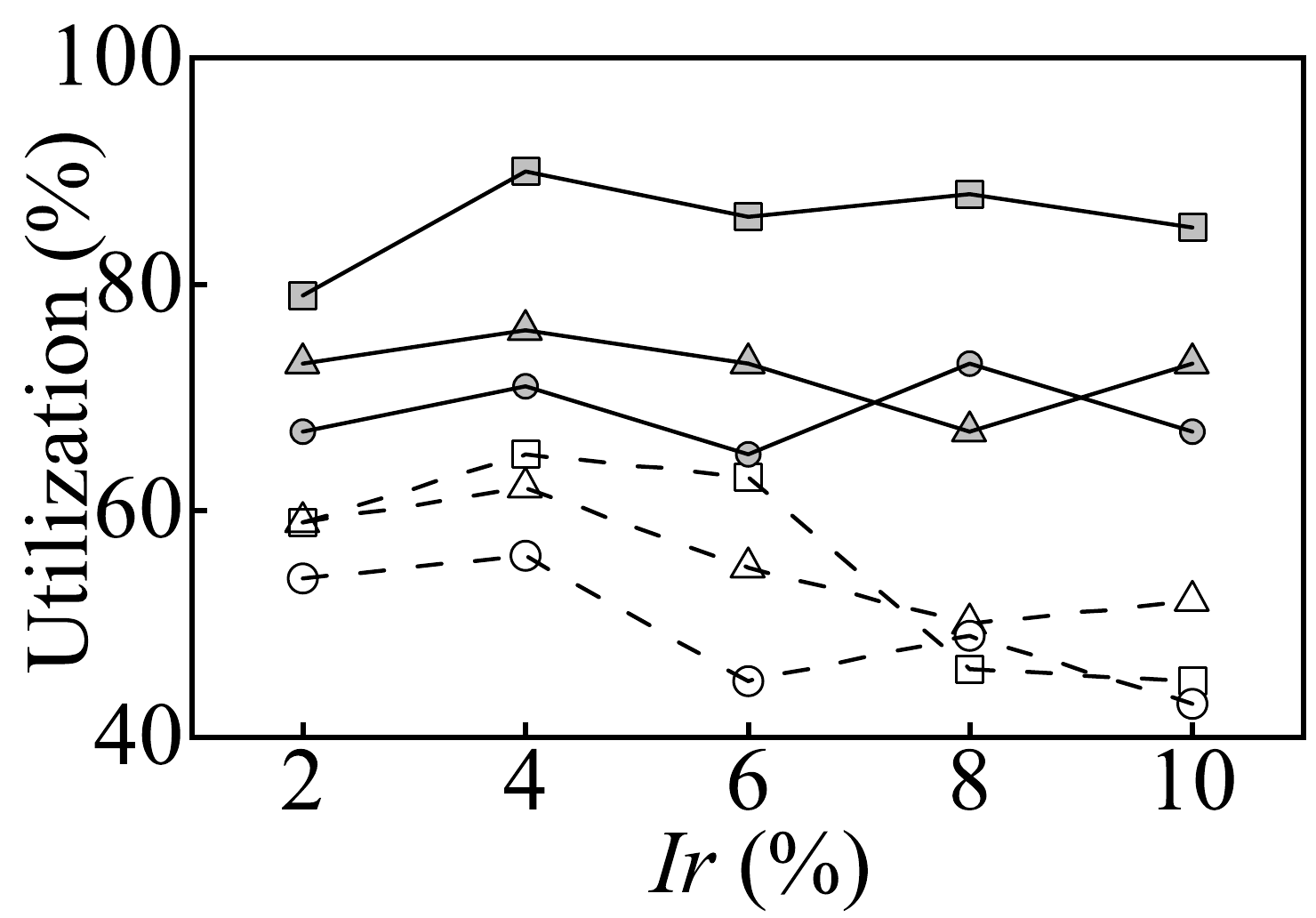}
    }

\vspace{-3mm}
    \caption{GPU utilization vs. query graph size $|V(Q)|$ and insertion rate $I_r$ (*ws = work stealing)}
    \vspace{-5mm}
    \label{fig:utilization}
\end{figure*}

\subsection{Ablation Study}
\label{sec:ablation}
Lastly, we conduct an ablation study to assess the efficacy of each individual technique.

\textbf{{Effect of Stealing Strategy on GPU Utilization.}} We first evaluate the impact of work stealing on GPU utilization. The results are shown in Figure~\ref{fig:utilization}, where ``ws” represents the work stealing optimization. In general, \textsf{GAMMA} with work stealing consistently achieves higher GPU utilization compared to \textsf{GAMMA} without work stealing (\textsf{GAMMA} w/o ws), with an average 17.5\% increase and a peak improvement of 33.8\%. Upon comparing the enhancements in utilization across different query sets, it becomes evident that denser queries exhibit a more modest improvement, owing to their comparatively reduced result sets and runtime. Consequently, this leads to a diminished disparity in cumulative execution time across warps. As query size and insertion rate increase, GPU utilization generally declines due to the expanding search space and workload, leading to larger disparities in cumulative execution time among warps. Additionally, as query size and insertion rate increase, the gap in utilization between schemes with and without work stealing progressively widens, demonstrating work stealing's effectiveness in balancing workloads among different warps and improving GPU utilization.

\begin{figure}[tb]

    \centering
    \includegraphics[width=0.42\textwidth]{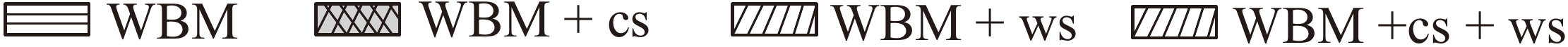}
    \vspace{0mm}

    \hspace{-4mm}
    \subfigure[Dense queries]{
    \includegraphics[width=0.16\textwidth]{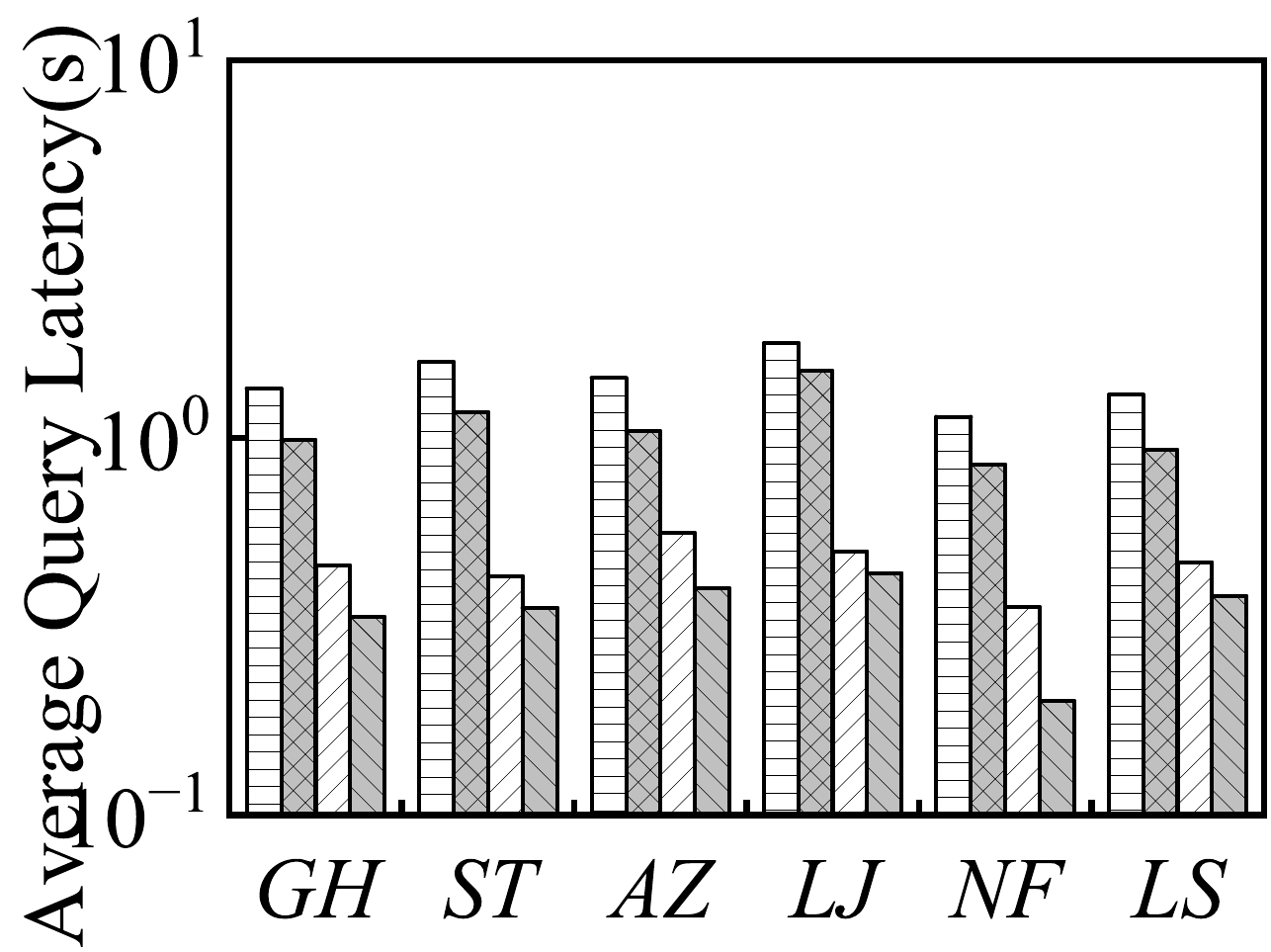}}
    \hspace{-4mm}
    \subfigure[Sparse queries]{
    \includegraphics[width=0.16\textwidth]{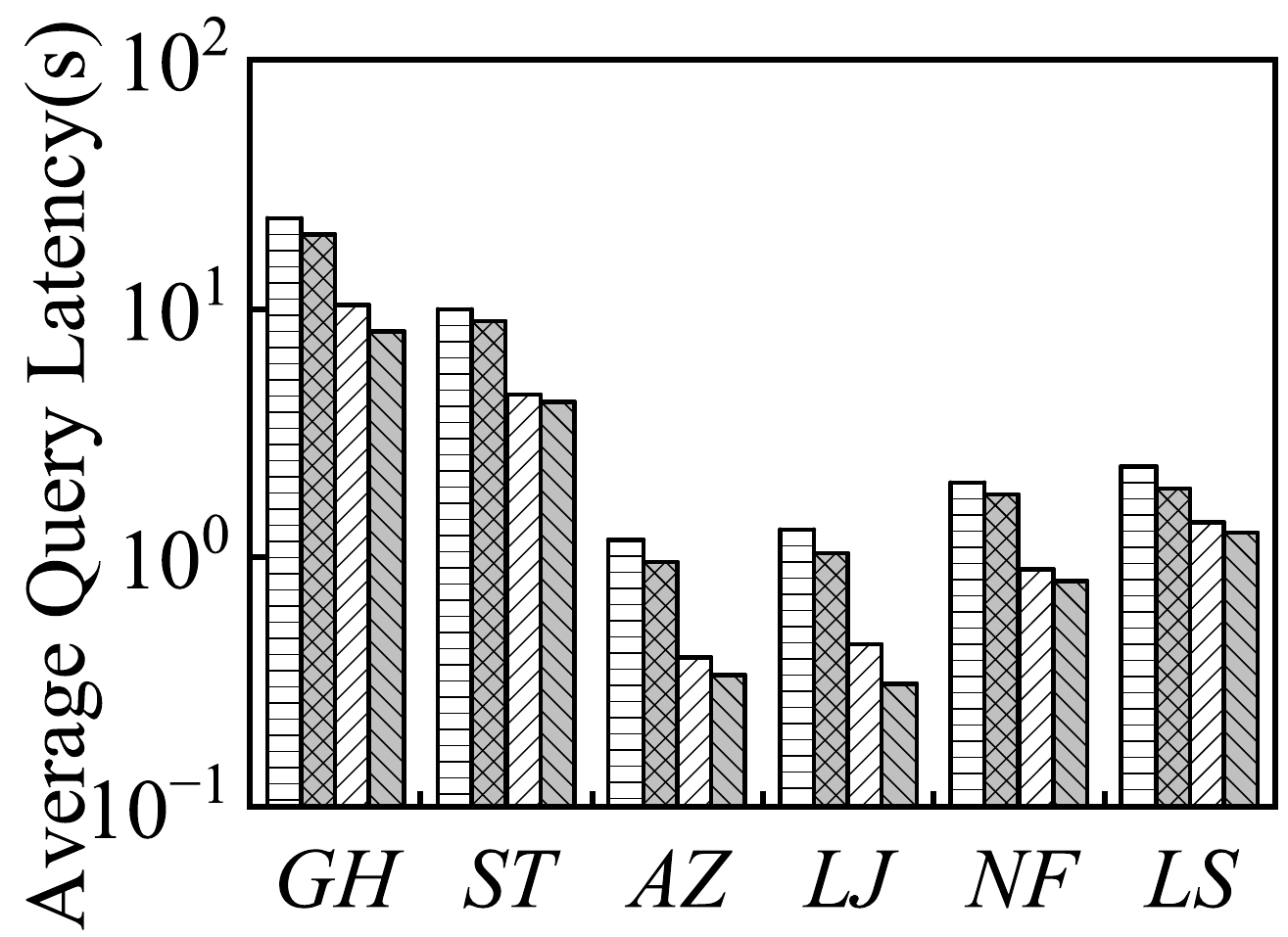}} 
    \hspace{-4mm}
    \subfigure[Tree queries]{
    \includegraphics[width=0.16\textwidth]{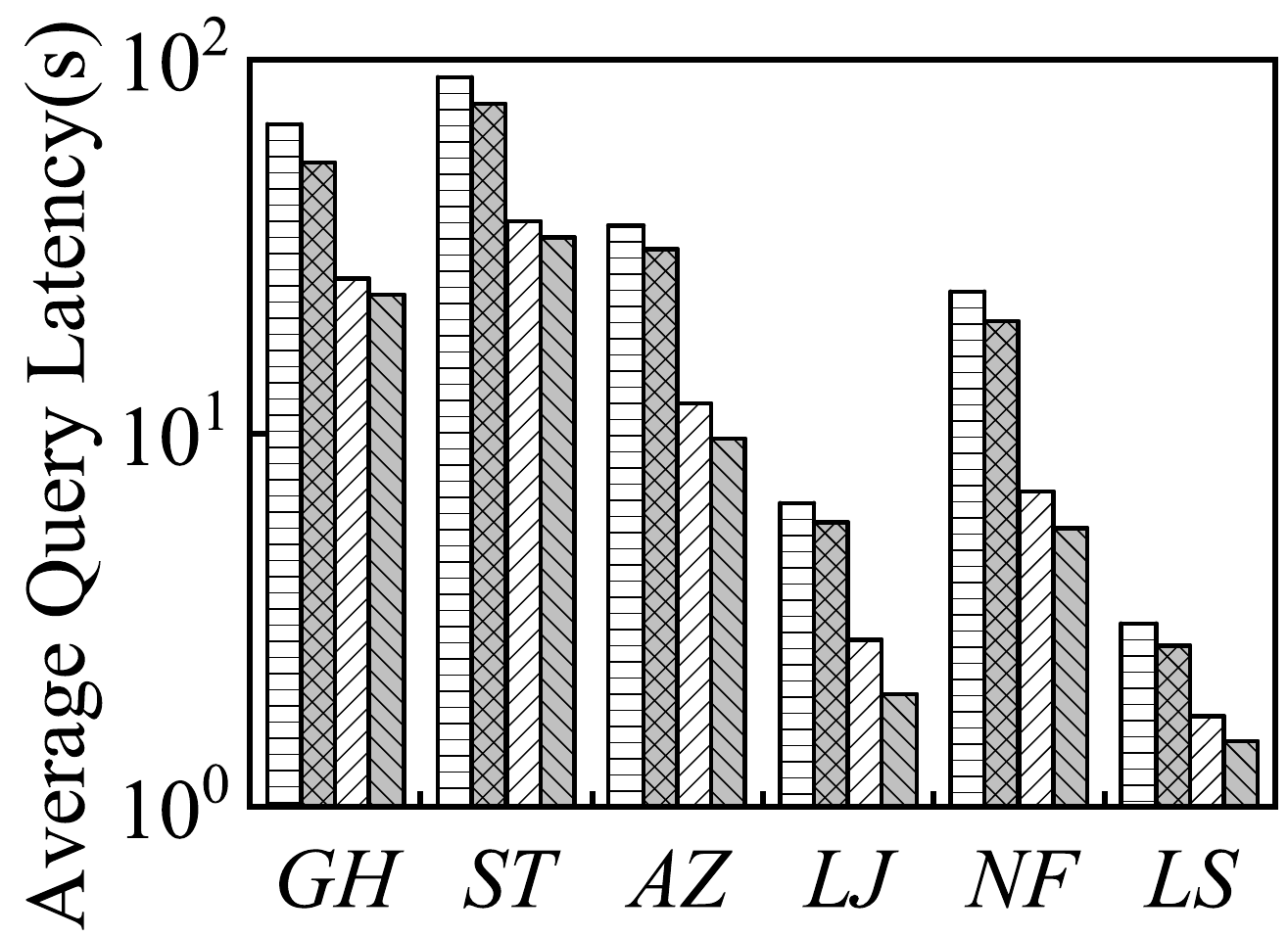}} 
    
    \hspace{-5mm}
    \vspace{-5mm}
    \caption{ {Ablation study}}
     \vspace{-6mm}
    \label{fig:abl}
\end{figure}

\textbf{{Effect of Stealing Strategy and Coalesced Search on Execution Time.}}
We further evaluate the influence of various techniques on performance. The results are reported in Figure \ref{fig:abl}, where ``cs" and ``ws" denote the coalesced search and work stealing, respectively.
The initial findings reveal that all the other implementations outperform WBM without optimizations, confirming the effectiveness of our proposed techniques. Notably, the load-balanced implementation (WBM+ws) executes faster than that with coalesced search (WBM+cs), underscoring the paramount importance of sophisticated load balancing techniques.
When comparing different query sets, we discern significantly higher speedup ratios for sparser queries, mainly due to their larger search space. Coalesced search effectively curtails this search space, resulting in noteworthy improvements in speedup. Consequently, the speedup ratios for sparse and tree queries are substantially greater in comparison to dense queries. Overall, the coalesced search achieves a speedup ranging from 1.1$\times$ to 1.9$\times$, and the work stealing delivers performance enhancements ranging from 1.2$\times$ to 6.4$\times$. In conclusion, our well-conceived optimization techniques significantly enhance performance.

\section{Conclusion}
\label{sec:conclusion}
This paper introduces \textsf{GAMMA}, an efficient parallel subgraph matching system tailored for batch-dynamic graphs. Our system harnesses a warp-centric parallel algorithm as its core, adeptly managing each update. To achieve balanced workloads among warps within a block, we implement a work stealing mechanism that effectively utilizes shared memory. Moreover, we integrate a coalesced search technique to mitigate redundant computations arising from automorphisms of subgraphs in the query graph. Lastly, we synergize these techniques with multiple other optimizations, culminating in a comprehensive bottom-up batch-dynamic subgraph matching system. Experiments conducted on four real-world datasets substantiate that our system surpasses state-of-the-art methods by a substantial margin. In the future, we envision expanding \textsf{GAMMA}'s capabilities to address more general subgraph matching challenges within batch-dynamic scenarios.

\balance

\section{Acknowledgement}

This was supported in part by the NSFC under Grants No. (U23A20296, 62025206, 62102351), Yongjiang Talent Introduction Programme (2022A-237-G). Xiangyu Ke is the corresponding author of the work.

\balance

\bibliographystyle{abbrv}
\bibliography{ref}

\end{document}